\documentclass[sigconf]{acmart}

\usepackage{graphicx}
\usepackage{subcaption}
\usepackage{amsmath}
\usepackage{enumitem}
\usepackage{url}   

\title{CarbonSet: A Dataset to Analyze Trends and Benchmark the Sustainability of CPUs and GPUs}

\newif\ifanonymous

\ifanonymous
  \author{Anonymous Authors}
\else
  \author{Jiajun Hu}
  \affiliation{
    \institution{Arizona State University}
    \city{Tempe}
    \state{AZ}
    \country{USA}
  }
  \email{jiajunh5@asu.edu}

  \author{Chetan Choppali Sudarshan}
  \affiliation{
    \institution{Arizona State University}
    \city{Tempe}
    \state{AZ}
    \country{USA}
  }
  \email{cchoppal@asu.edu}

  \author{Vidya A. Chhabria}
  \affiliation{
    \institution{Arizona State University}
    \city{Tempe}
    \state{AZ}
    \country{USA}
  }
  \email{vachhabr@asu.edu}

    \author{Aman Arora}
  \affiliation{
    \institution{Arizona State University}
    \city{Tempe}
    \state{AZ}
    \country{USA}
  }
  \email{aman.kbm@asu.edu}

\fi

\begin{document}

\begin{abstract}
Over the years, the chip industry has consistently developed high-performance processors to address the increasing demands 
across diverse applications. However, the rapid expansion of chip production has significantly increased carbon emissions, raising critical concerns about environmental sustainability. While researchers have previously modeled the carbon footprint (CFP) at both system and processor levels, a holistic analysis of sustainability trends encompassing the entire chip lifecycle remains lacking.
This paper presents CarbonSet, a comprehensive dataset integrating sustainability and performance metrics for CPUs and GPUs over the past decade. CarbonSet aims to benchmark and assess the design of next-generation processors. Leveraging this dataset, we conducted detailed analysis of flagship processors' sustainability trends over the last decade. This paper further highlights that modern processors are not yet sustainably designed, with total carbon emissions increasing more than 50$\times$ in the past three years due to the surging demand driven by the AI boom.
Power efficiency remains a significant concern, while advanced process nodes pose new challenges requiring to effectively amortize the dramatically increased manufacturing carbon emissions.

\end{abstract}




\maketitle

\section{Introduction}
\label{sec:intro}
\label{sec:intro_motivation}
\noindent
The information and communication technology (ICT) sector contributes 2.1\%–3.9\% of global greenhouse gas emissions with emissions projected to rise~\cite{chasing_carbon}. Emissions mainly arise from chip manufacturing, design, and packaging (embodied carbon) and energy consumption of daily operation (operational carbon). Designing more sustainable processors for both manufacturing and daily use is vital.
Recent works include tools for modeling system-level embodied CFP \cite{ACT} and further chip-level modeling for general processors and reconfigurable systems \cite{eco-chip, greenFPGA}. Some studies also propose new sustainability-focused metrics for architecture design space exploration \cite{metrics, metrics_acm}.  
Historical insights of trend analysis and projections of processor metrics have long been integral to the computing industry. For example, Moore's Law—central to driving semiconductor innovation for over 50 years—was understood by curating and analyzing data from numerous chips \cite{itrs}. Similarly, analyzing trends in CFPs of modern processors (CPUs and GPUs) can raise awareness of their environmental impact, identify areas for intervention, and guide the design of sustainable chips.

\subsection{Goals}
\label{sec:intro_goals}  
\noindent
This paper seeks to create actionable insights into processor sustainability while enabling benchmarking of existing processors based on their CFP. We aim to raise semiconductor community awareness about the growing sustainability challenges computing technologies pose. We curate a dataset of CPUs and GPUs over the last decade, analyzing their CFP across lifecycle stages for both datacenter and desktop series. We aim to answer the following key research questions:
\begin{itemize}
    \item How has the CFP of flagship GPUs and CPUs evolved?  
    \item Does the increased performance of chips justify higher CFP?  
    \item For flagship processors, which type of CFP dominates?
    \cite{andrew-chien-sigarch-blog, david-brooks-sigarch-blog}
    \item Has the AI boom impacted the CFP of processors?  
    \item Is the chip price (\$) a reliable proxy for its Embodied CFP?  
    \item What processor lifetimes effectively amortize Embodied CFP? 
    \item Are chiplet-based processors always more sustainable than monolithic processors?
\end{itemize}

\noindent
Answers to these questions help identify key trends and sustainability challenges and also enable more environmentally conscious decision-making in chip design, manufacturing, and lifecycle management.
 
\subsection{Overview of Our Contributions}
\label{sec:intro_overview}
\begin{figure}[t]
    \centering
    \includegraphics[width=1.0\linewidth]{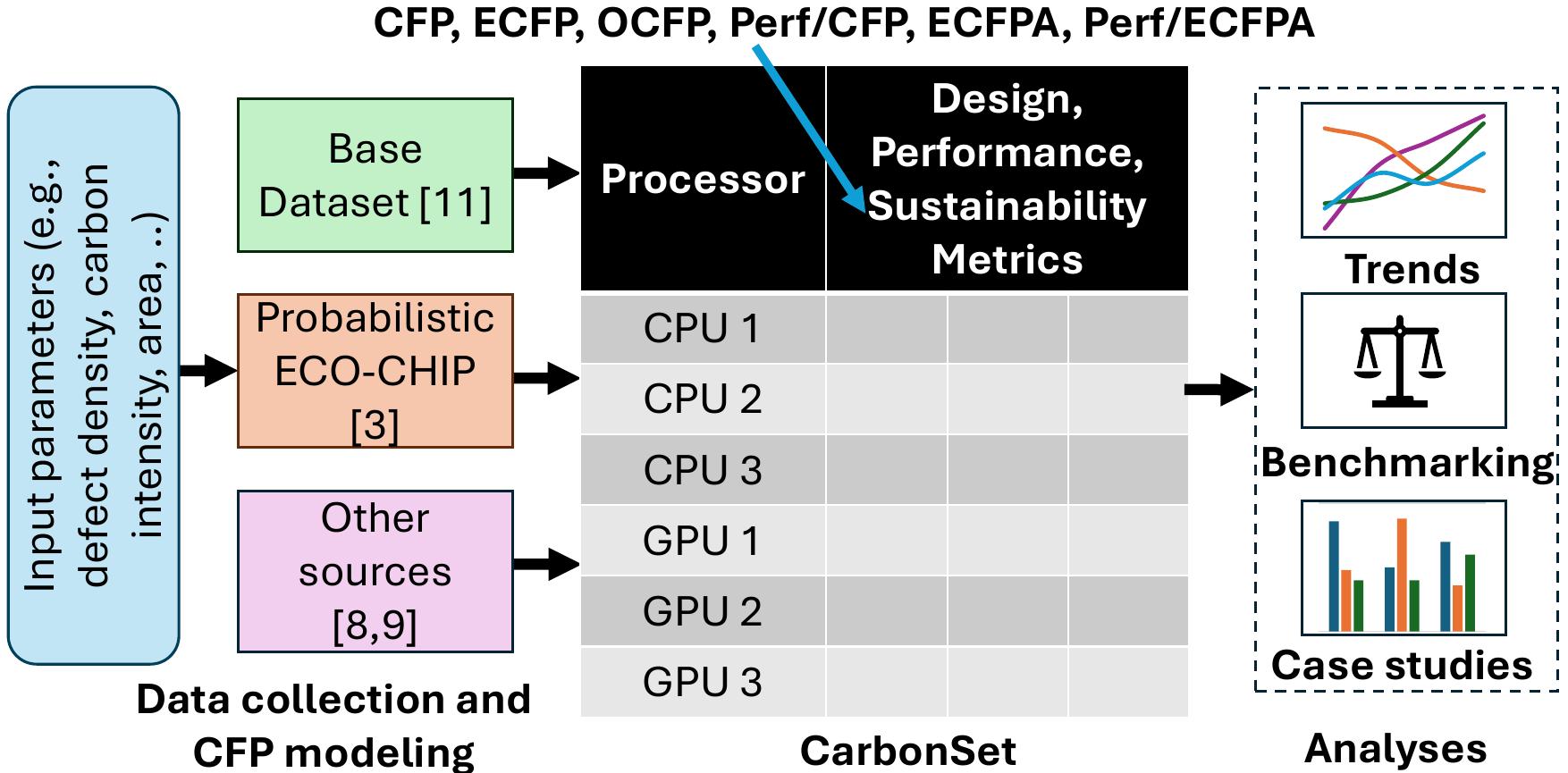}
    \caption{CarbonSet contains sustainability-related metrics for multiple CPUs and GPUs derived from probabilistic carbon modeling.}
    \vspace{-3mm}
    \label{fig:carbonset_overview}
\end{figure}

\noindent
Figure. \ref{fig:carbonset_overview} shows an overview of CarbonSet, including data for desktop and datacenter CPUs and GPUs across multiple metrics - {\it design metrics} (chip area, technology node, transistors, TDP), {\it performance metrics} (throughput, OpenCL \cite{Geekbench}, Passmark \cite{PassMark}), and \textit{sustainability metrics} (Embodied CFP, Operation CFP, total CFP). It also offers composite metrics like Performance per CFP and ECFP per unit area (ECFPA) for further tradeoff-based trend analysis.

Estimating CFP across a processor’s lifecycle is challenging due to uncertainties in parameters such as manufacturing defect density, energy source carbon intensity, utilization patterns, and lifetimes. 
CarbonSet addresses this issue by extending the ECO-CHIP framework \cite{eco-chip} to generate ranges of CFP values instead of a single CFP value. This range of CFP values are evaluated using several probabilistic modeled parameters, which are derived from real-world data and have more practical significance.

From our analysis, we find that flagship GPUs and CPUs are still far away from achieving sustainable design. Although operational CFP (OCFP) still dominates throughout the lifecycle, the proportion of embodied CFP (ECFP) is also increasing due to the advancement of the process nodes.
Moreover, in our estimation, given the performance per unit CFP increased by over 100$\times$ in some cases, the dramatically increased chip shipments, driven by the AI boom, have eventually led to 50$\times$ more total CFP.

To the best of our knowledge, CarbonSet is the first work for processor sustainability evaluation, containing more than 1000 processors, out of which 45\% are GPUs and 55\% are CPUs. This includes monolithic GPUs, monolithic CPUs, and recent chiplet-based CPUs as well. By leveraging CarbonSet, a Moore's law-like trend analysis is performed to understand how processor sustainability has evolved, which could also be used for `pathfinding' studies for future architecture design. With more detailed processor specifications and modeling, CarbonSet could also serve as an industry processor sustainability evaluation norm. Our contributions in this work include the following:
\begin{itemize}[leftmargin=*,itemsep=0pt,topsep=0pt]
\item {\textbf{\underline{\textit{Comprehensive dataset curation:}}}} We curate a dataset of CPUs and GPUs across desktop and datacenters from multiple vendors, spanning the last decade. This dataset includes detailed design, performance, and sustainability metrics. CFP is evaluated across all stages of the processor lifecycle - design, manufacturing, use, and end-of-life. 
The complete dataset is available at~\cite{carbonset_github}. 

\item {\textbf{\underline{\textit{Probabilistic CFP modeling:}}}} We extend ECO-CHIP to generate ranges of CFP values for each processor based on probabilistic modeled key parameters, including defect density, carbon intensity, energy per area and gas per area. This mitigates the challenges arising from the uncertainties across a processor's lifecycle, resulting in practical ranges of CFP values instead of specific values, which are difficult to validate.

\item {\textbf{\underline{\textit{Sustainability analysis:}}}} We track the sustainability trends of flagship processors, by examining metrics such as ECFP, OCFP, performance per CFP, and ECFP per area. Such an analysis needs significant effort in selecting the most representative flagship processors across generations while ensuring the availability of sufficient performance benchmark data for each selected one. 

\item {\textbf{\underline{\textit{In-depth case studies:}}}} We also conduct several case studies to address the aforementioned research questions to better understand how external factors affect processor sustainability such as the impact of processor lifetime on ECFP amortization, the increasing shipment demands driven by the AI boom, increased manufacturing costs due to advanced process nodes, and the sustainability evaluation of contemporary chiplet architectures.

\end{itemize}

\vspace{-3mm}

\section{Related Work} \label{sec:related_work}

\noindent
\textbf{\underline{\textit{Datasets}}}:
Researchers have analyzed processor design trends using area, power and performance metrics. For example, \cite{dataset} reviewed processor design trends over the last two decades discussing the major driving factors of performance increases
Existing datasets, however, do not provide sustainability data for the various processors, which precludes the analysis of trends and the benchmarking of processors from a sustainability standpoint.

\noindent
\textbf{\underline{\textit{Carbon modeling}}}:
Driven by the rapidly growing carbon emissions due to the increased number of large-scale datacenters \cite{chasing_carbon, ml_carbon, sustainableAI}, researchers have proposed various CFP modeling frameworks.
These range from general first-order estimation \cite{first_order} to comprehensive system-level modeling (ACT) \cite{ACT} and support for heterogeneous integration (chiplet-based) chips (ECO-CHIP) \cite{eco-chip}.
Inspired by \cite{bhagavathula:understanding:2024}, We updated ECO-CHIP to generate probabilistic modeled CFP data and then use it to complete CarbonSet.

\noindent
\textbf{\underline{\textit{Metrics}}}:
Chip designers and architectures have used traditional metrics such as chip area, frequency, power, performance-per-watt, and area-delay-product to benchmark chips across generations. 
However, such metrics fail to capture sustainability-related tradeoffs.
While ECFP and OCFP quantify carbon emissions, new metrics like CDP (Carbon Delay Product), CEP (Carbon Energy Product) \cite{ACT}, performance per unit CFP (Perf-SI), and CFP per billion transistors \cite{metrics} offer improved tradeoff evaluation. Our dataset includes metrics like CFP, ECFP, OCFP, performance per CFP,
embodied CFP per area (ECFPA), and performance per ECFPA.
\vspace{-3mm}
\section{Modeling and Dataset Contents}
\label{sec:Carbonset Modeling}
\subsection{Carbon Footprint Modeling}

\noindent
\textbf{\underline{\textit{Prior carbon modeling:}}}
In our dataset, CFP is used as the primary metric to estimate the sustainability of a chip. We use ECO-CHIP \cite{eco-chip} to evaluate the CFP in all stages of the chip lifecycle, design, manufacturing, use, and end of life.
As modeled in ECO-CHIP, total CFP is the sum of ECFP and OCFP. ECFP includes the CFP spent during chip design, manufacturing and packaging. 
The manufacturing CFP depends on yield and the CFP per unit area (CFPA) of the manufacturing process.
Yield is influenced by die area and can be calculated using the equation:
\begin{equation}
\label{eq:yield}
    Y = \left ( 1 + \frac{A_\text{die} \times D_0}{\alpha} \right ) ^ {-\alpha} 
\end{equation}
where $A_{\text{die}}$ is die area, $\alpha$ is the clustering parameter, and $D_0$ is defect density of current process node, and CFPA is given by~\cite{ACT,eco-chip}:
\begin{equation}
\label{eq:cfpa}
    \text{CFPA}  = \frac{(C_{\text{mfg, src}} \times \text{EPA}  + C_{\text{gas}} + C_{\text{material}})}{Y}  
\end{equation}

\noindent
where $C_{\text{mfg, src}}$ is the carbon intensity, and it mainly depends on the source of energy for fabrication, EPA is the energy consumed per unit area (kWh per cm$^2$), $C_{\text{gas}}$ is a function of gas per unit area (GPA) emissions usually expressed in kgs of CO$_2$eq 
per cm$^2$, and $C_{\text{material}}$ is the emission from procuring materials expressed in kgs of CO$_2$eq 
per cm$^2$, and $Y$ is the yield given in Eq.~\ref{eq:yield}.

OCFP is modeled as the carbon intensity of the energy source times the energy spent during operation. The latter is determined by using the TDP and the lifetime scaled by the time during which the chip is in idle state (idle time).

\noindent

\textbf{\underline{\textit{Probabilistic carbon modeling}}}:
CFP values are difficult to validate due to the inherent uncertainties of the chip's lifecycle analysis.
Therefore, inspired by \cite{bhagavathula:understanding:2024}, we enhance ECO-CHIP to produce a range of CFP values by modeling multiple input parameters as probability distributions. The resultant CFP distribution now represents the viable range of the chip's CFP values instead of a single value.

\begin{figure}[t]
    \centering
    \vspace{-3mm}
    \includegraphics[width=0.9\linewidth]{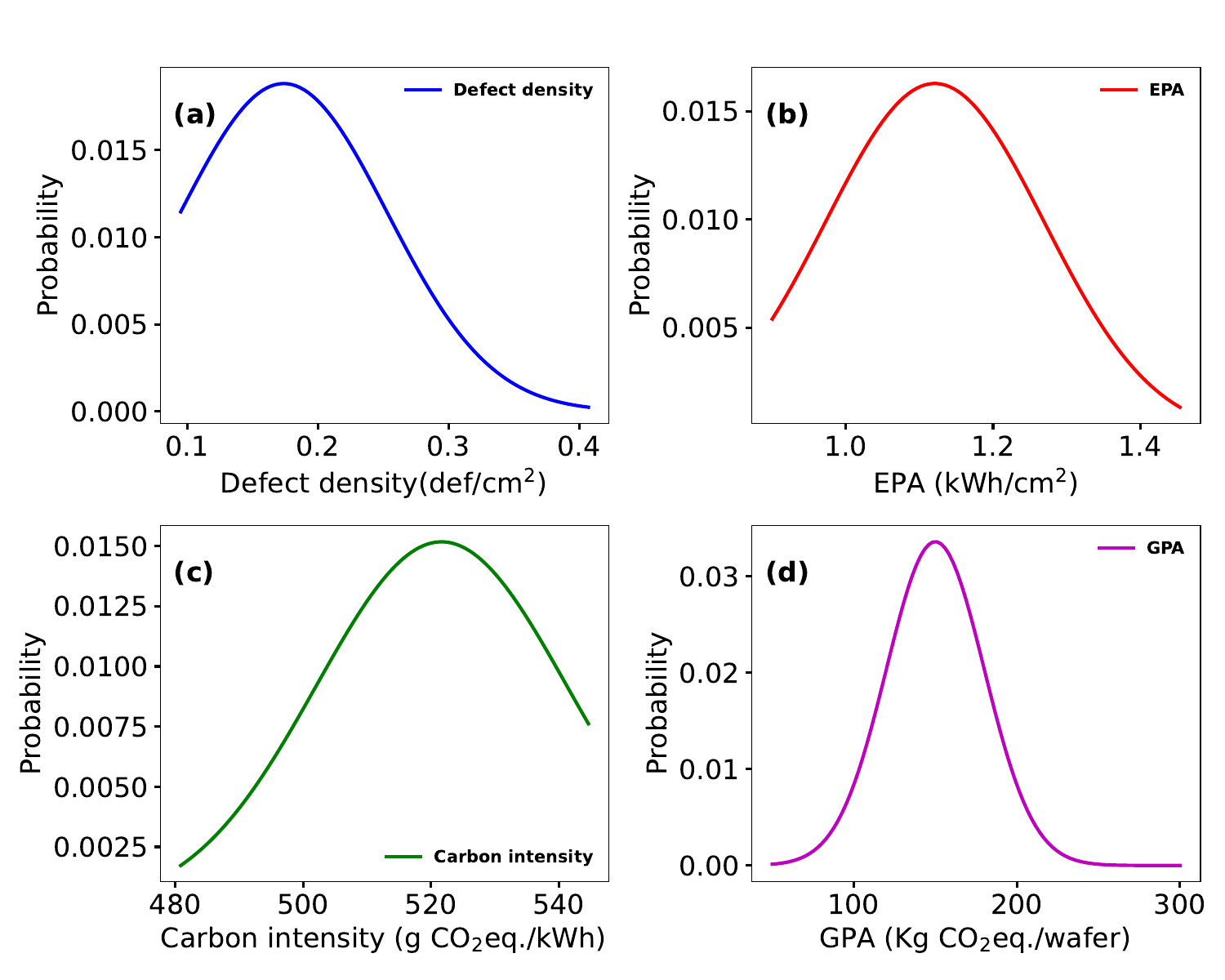}
    \caption{Distributions for (a) Defect density(10nm) ~\cite{def-den} (b) Energy-Per-unit Area(EPA)(10nm) ~\cite{epa-tsmc} (c) Carbon intensity ~\cite{ci-world} (d) Gas-Per-unit Area(GPA) ~\cite{gpa-imec} }
    \label{fig:uncertainity_params}
    \vspace{-5mm}
\end{figure}

\begin{figure*}[h]
    \centering
    \includegraphics[width=1.0\linewidth]{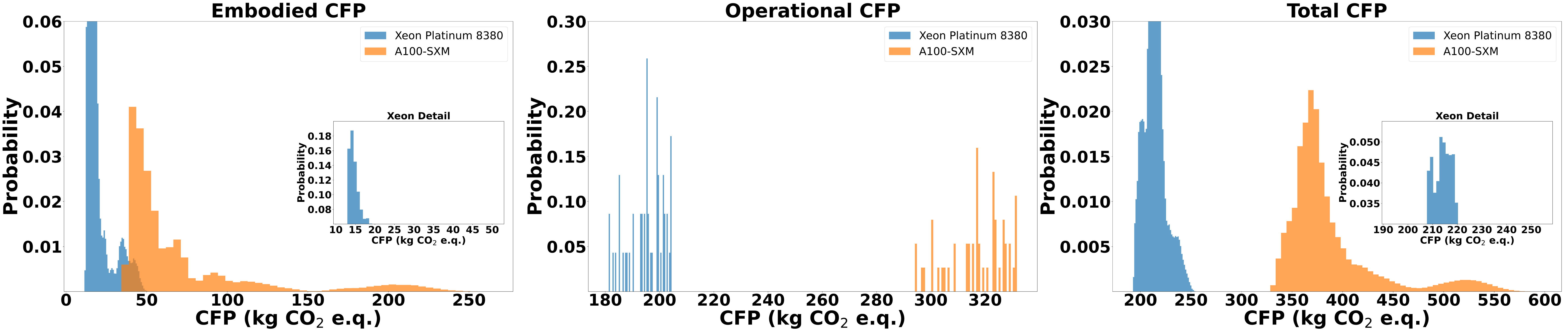}
    \caption{CFP distributions for two flagship processors from our dataset obtained from enhanced ECO-CHIP, by varying defect density, EPA, GPA, carbon intensity, lifetime, and idle time.}
    \label{fig:range_chart}
    \vspace{-5mm}
\end{figure*}

Figure.~\ref{fig:uncertainity_params} illustrates the probability density function of four input parameters: defect density, EPA, carbon intensity, and GPA. Defect density, EPA, and carbon intensity are modeled using Kernel Density Estimation (KDE), while GPA follows a Gaussian distribution.  
The distributions of defect density and EPA are derived from 10nm TSMC reports ~\cite{def-den, epa-tsmc}. Carbon intensity is modeled by analyzing global trends over the past 24 years from ~\cite{ci-world}. The GPA is sourced from IMEC ~\cite{gpa-imec}, which examines variations in greenhouse gas emissions at fabrication facilities.  For OCFP model, a fixed three-year lifetime with 60\% idle time is assumed.

Using enhanced ECO-CHIP, we perform Monte Carlo simulations with 10,000 samples per input parameter to model CFP variability across processors. Figure~\ref{fig:range_chart} presents the resulting CFP distributions for the A100-SXM and Xeon Platinum 8380. By incorporating parameter uncertainty, ECO-CHIP supports what-if analyses for carbon-aware design decisions. Overlapping distribution regions indicate that the Xeon is not consistently more sustainable than the A100. For consistency, we use the mean CFP from the 10,000 simulations as the final value for each processor.


\subsection{Dataset Contents}
\label{sec:metrics}
\noindent
CarbonSet builds on a dataset of over 1,000 CPUs and GPUs from vendors such as Intel, NVIDIA, and AMD, covering both datacenter and desktop series. While the base dataset supports trend analysis using conventional metrics, CarbonSet adds a probabilistic model that estimates a CFP range for each processor, using the mean as a representative value. It includes 30 chiplet-based CPUs, assuming equal die area distribution and a uniform process node. Users can modify these assumptions as more vendor architectural details become available. 

Due to the complicated system/software configurations, it is difficult to establish a single-chip performance benchmark applicable to both desktop and datacenter CPUs. \textit{SPEC} \cite{spec} is a widely accepted CPU performance benchmark; however, comparing scores from different \textit{SPEC} versions (e.g., \textit{SPEC-2017} and \textit{SPEC-2006}) is not valid as each \textit{SPEC} version uses distinct evaluation suites. Therefore, for CPU benchmarking, we use the highest single-chip scores from \textit{Geekbench} for desktops and \textit{Passmark} for datacenter series, respectively \cite{Geekbench, PassMark}. 

For GPUs, the performance metrics listed in the vendor specifications are rarely achieved in practice. Also, due to varying precision support and the introduction of Tensor Cores, comparing peak metrics across generations is unreliable (e.g., comparing Float16 performance of the A100 with Float32 performance of the P100). Hence, we use the \textit{OpenCL score} from GeekBench as a performance metric of all GPUs \cite{Geekbench}. To evaluate performance-sustainability tradeoffs, we include data for the following metrics inspired by \cite{metrics}:
\begin{itemize}[leftmargin=*,itemsep=0pt,topsep=0pt]
    \item \textbf{\underline{\textit{Performance per unit CFP:}}} 
    This metric measures the performance gain per unit of CFP, which is an overall evaluation factor of the balance between sustainability and performance comprising both the embodied and operational CFP. 
    
    \item \textbf{\underline{\textit{Embodied CFP per area (ECFPA):}}}
    ECFPA assesses the $CO_2$ density per unit chip area, which is related to the chip manufacturing process node. Generally, an advanced process node has a higher CO$_2$ emission due to complex lithography processes and lower yields. However, this factor can also be significantly influenced by chip architecture; for example, chiplet-based designs may reduce the overall ECFPA\cite{eco-chip}. An architecture with a complicated design process may also increase the design carbon. This metric, therefore, assesses the chip's sustainability and scalability in the design and manufacturing stages.

    \item \textbf{\underline{\textit{Performance per ECFPA:}}}
   This metric assesses performance gain relative to ECFP and helps to evaluate the performance scalability of the die area for mass production. A chip with high ECFPA but only marginal performance is unlikely to meet sustainability standards. Ideal processors balance low ECFPA with high performance to meet demands sustainably.
    
\end{itemize}

\vspace{-3mm}
\section{Data Analysis and Trends} 
\label{sec:results}
\noindent
To showcase the trends in CFP, NVIDIA GPUs and Intel CPUs are selected for their leading roles in datacenter and desktop series, which also offer extensive documentation and architectural specifications. Since most hardware vendors diversify product lines by binning and only flagship chips with the largest die and highest TDP are the most complete designs, our dataset only selects those most representative flagship models across both desktop and datacenter series. 

\vspace{-3mm}

\subsection{Sustainability Trends in GPUs}
\begin{figure*}[h]
\centering
\begin{subfigure}{0.49\textwidth}
    \includegraphics[width=\textwidth]{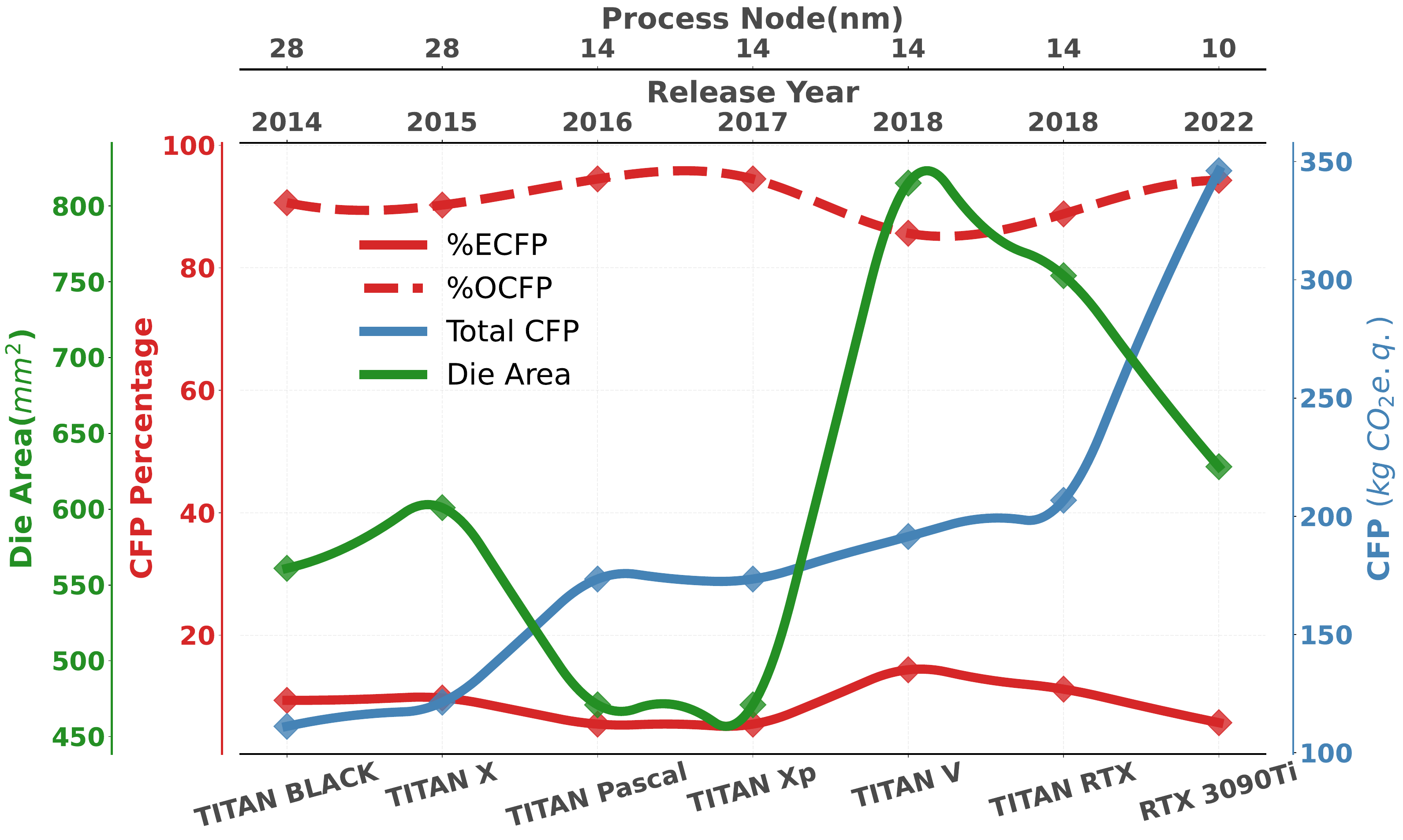}
\end{subfigure}
\hfill
\begin{subfigure}{0.46\textwidth}
    \includegraphics[width=\textwidth]{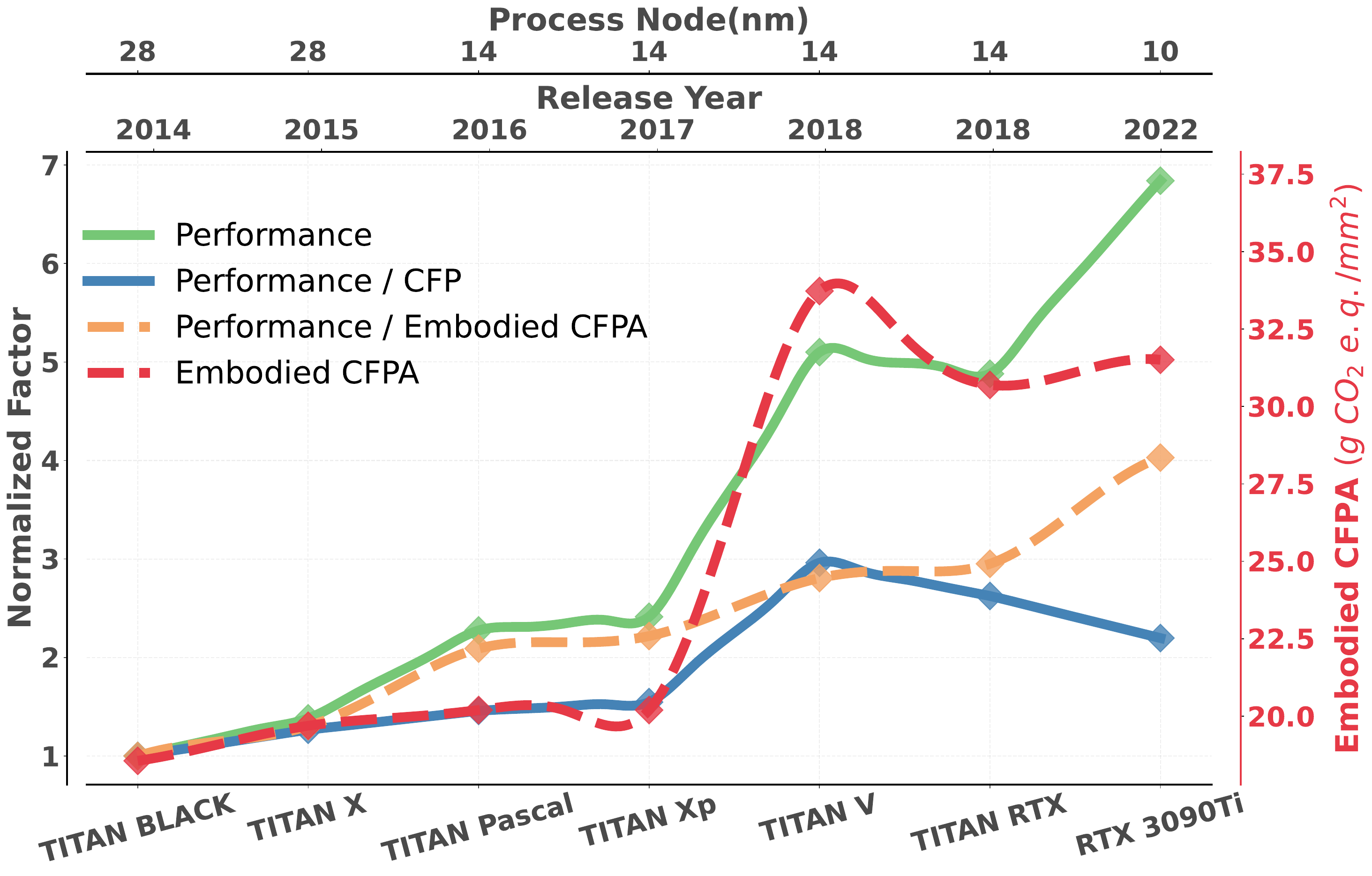}
\end{subfigure}

\begin{subfigure}{0.49\textwidth}
    \includegraphics[width=\textwidth]{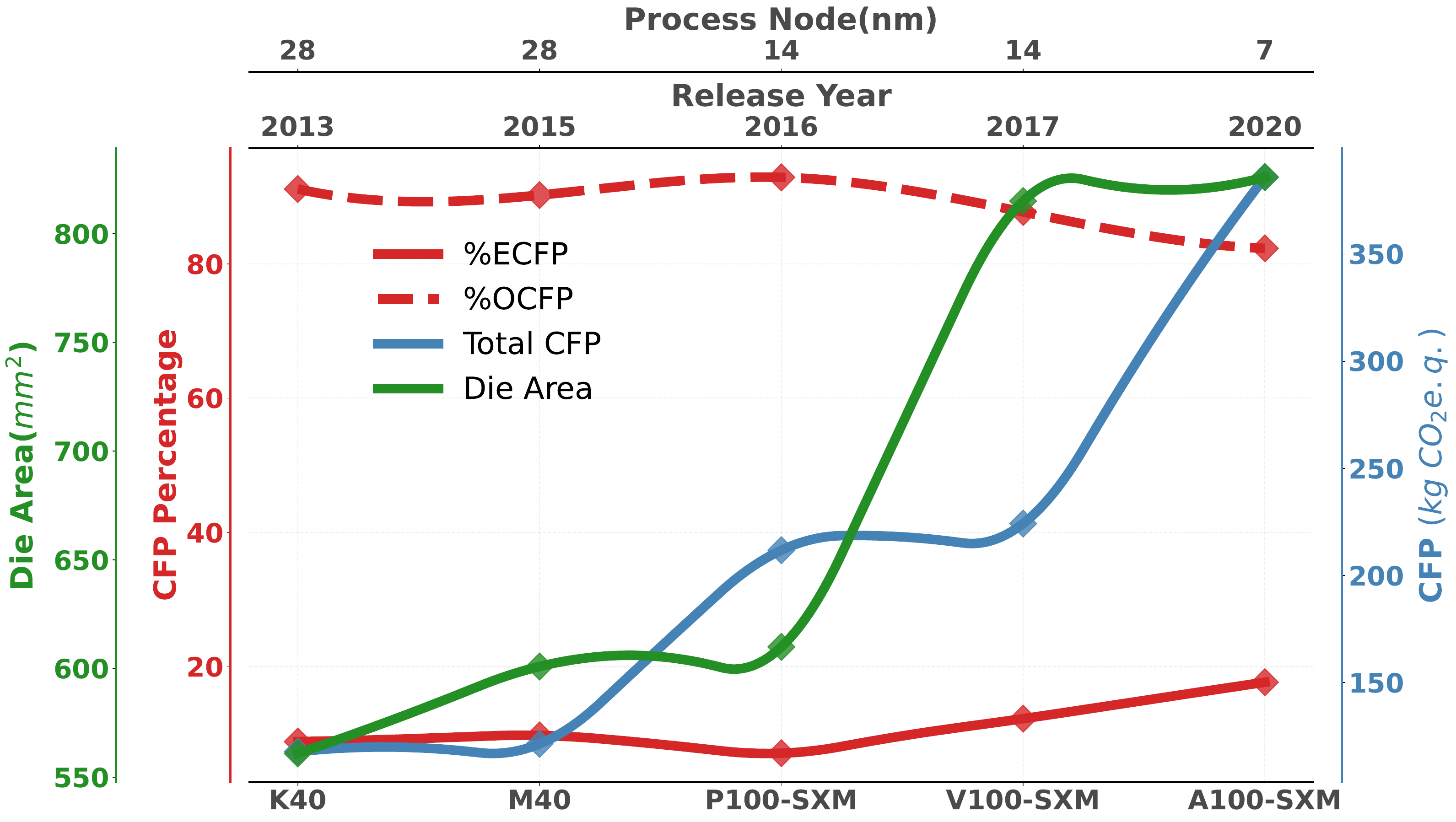}
\end{subfigure}
\hfill
\begin{subfigure}{0.46\textwidth}
    \includegraphics[width=\textwidth]{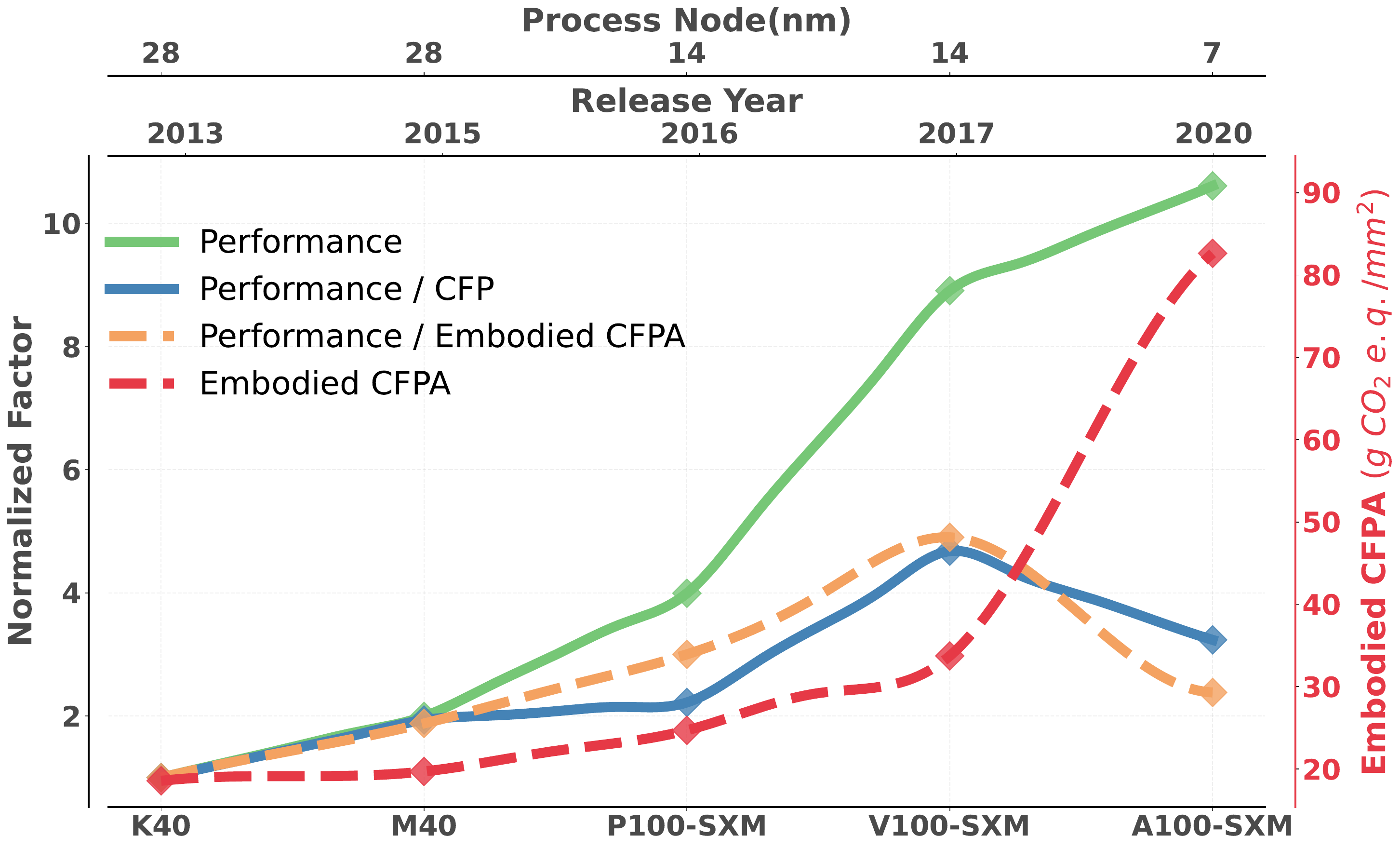}
\end{subfigure}

\caption{Sustainability trends in flagship desktop (top) and datacenter (bottom) GPUs from NVIDIA}
\vspace{-5mm}
\label{fig:gpu_sustainability}
\end{figure*}

\noindent
The left two plots in Figure. \ref{fig:gpu_sustainability}
show various design and sustainability metrics for flagship desktop and datacenter GPUs from NVIDIA in the last decade, respectively.
Both the datacenter and desktop GPU series exhibit an increasing trend in single-chip total CFP and OCFP, consistently constituting its majority. 

In the desktop series, the die area shows significant fluctuations, particularly at the 14nm node from 2016 to 2018, indicating continuous optimization of architecture design during this period. In contrast, datacenter GPUs have consistently increased area over generations. While the die area may fluctuate significantly within the same process node, the total CFP remains stable. For example, in 14nm, TITAN V has significantly increased die size but has a similar CFP to the previous 2 generations. Also, P100 incurs significantly larger CFP but a slightly varied die area. This shows that the ECFP led by the process node and die area contributes much less CFP than OCFP which is dominated by the chip TDP.

The right two plots in Figure. \ref{fig:gpu_sustainability}
show various performance and sustainability-related metrics for flagship desktop and datacenter GPUs from NVIDIA in the last decade, respectively. The performance metrics are described in Section~\ref{sec:metrics}. While overall GPU performance continues to improve, the architectural trade-offs and priorities across generations differ significantly when evaluated through a combined perspective of sustainability metrics.
Specifically in the desktop series, comparing to TITAN V, TITAN RTX has a similar performance and a significantly smaller ECFPA. However, this does not result in an expected improvement in Performance/CFP nor Performance/ECFPA. This implies that, while a chip's (e.g., TITAN RTX's) architecture is optimized for large-scale manufacturing and incurs less ECFP, it may incur much higher OCFP and fail to achieve a greater performance-sustainability balance than previous generations (e.g., TITAN V). 
Still in the desktop series, the performance/CFP  continues to drop after 2018, even with a node advanced from 14nm to 10nm; this highlights that the performance improvements in recent desktop GPUs are being achieved at the expense of increased CFP.

Compared to desktop GPUs, datacenter series appear even more extreme in the pursuit of performance, with models A100-SXM and V100-SXM achieving significantly higher performance with a boom of CFP, specifically the dropping Performance/CFP and Performance/Embodied CFPA.
Although A100 achieves a performance boost of less than 2$\times$ compared to V100, it comes at the cost of a dramatically increased ECFPA (over 4$\times$). In contrast, V100 demonstrates significantly lower ECFPA with moderate performance. As a result, V100 achieves around 2$\times$ performance/CFP than A100.

In summary, GPUs generally prioritize performance over sustainability, with datacenter GPUs being particularly extreme. Moreover, the comparison of TITAN V and TITAN RTX shows that the OCFP may be one of the major setbacks preventing achieving a better performance-sustainability balance.
\vspace{-4mm}
\subsection{Sustainability Trends in CPUs}
\begin{figure*}[h]
\centering
\begin{subfigure}{0.49\textwidth}
    \includegraphics[width=\textwidth]{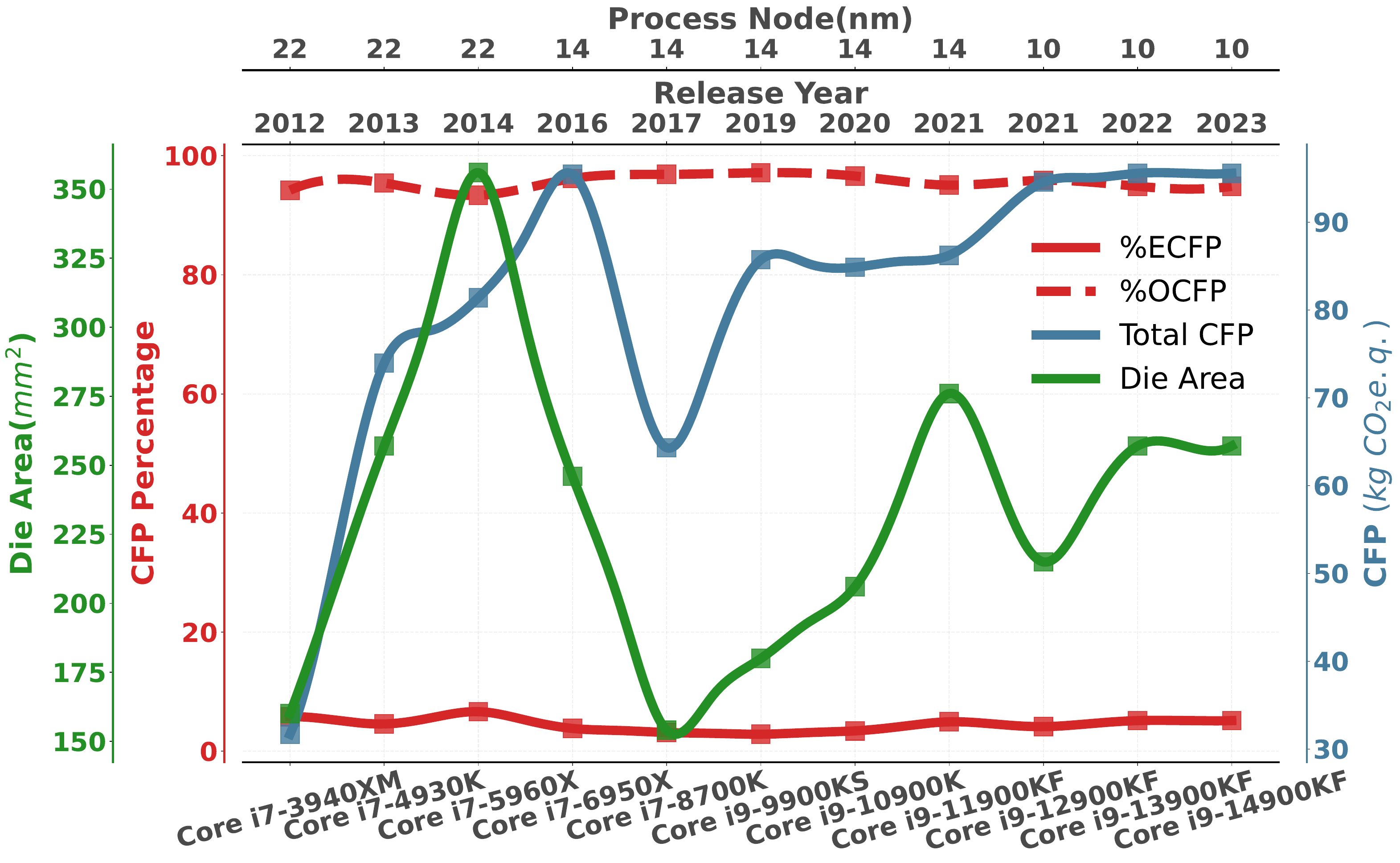}
    \label{fig:desktop_CPU_chip_metrics}
\end{subfigure}
\hfill
\begin{subfigure}{0.47\textwidth}
    \includegraphics[width=\textwidth]{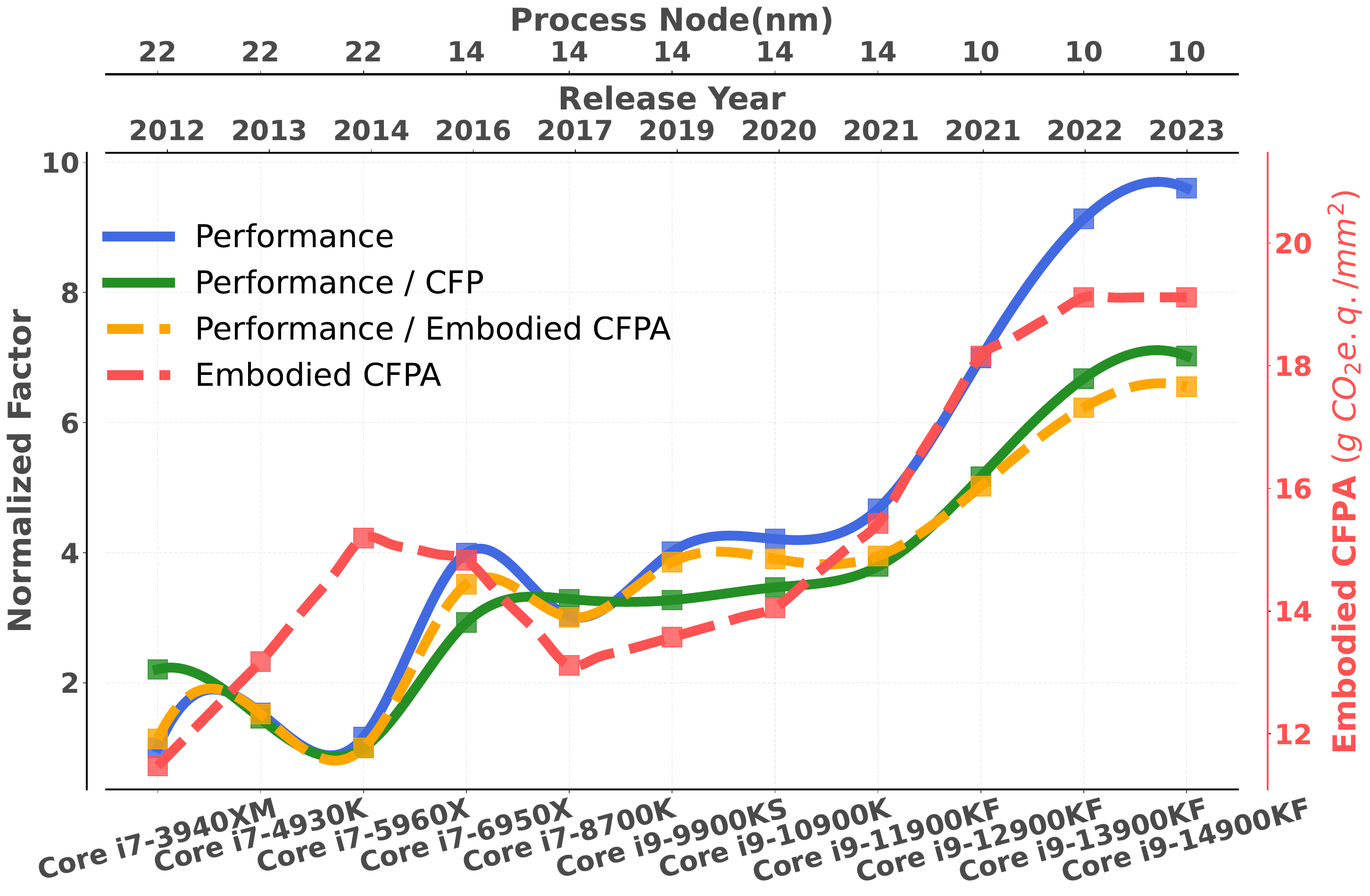}
    \label{fig:desktop_CPU_performance_metrics}
\end{subfigure}

\begin{subfigure}{0.49\textwidth}
    \includegraphics[width=\textwidth]{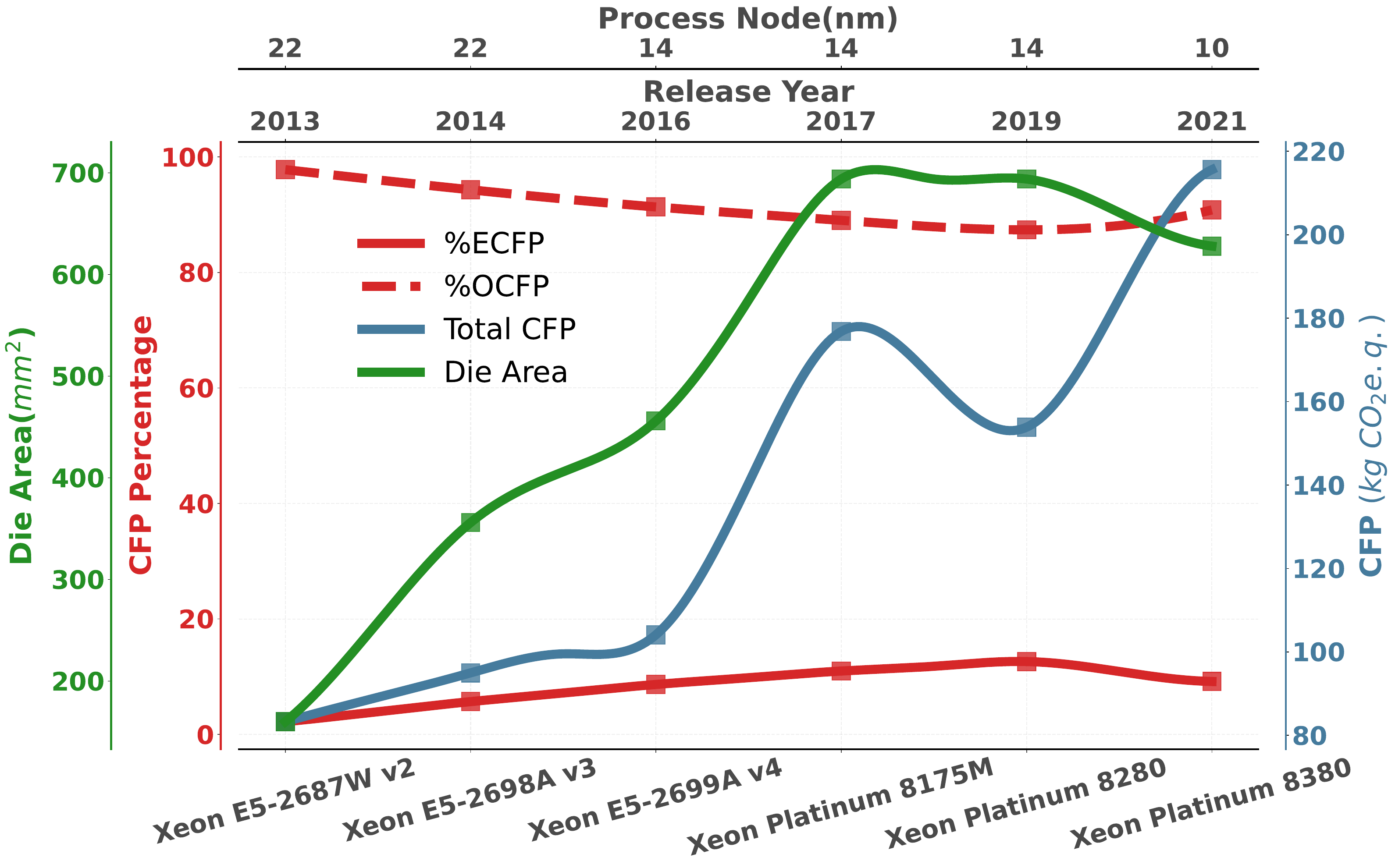}
    \label{fig:server_CPU_chip_metrics}
\end{subfigure}
\hfill
\begin{subfigure}{0.47\textwidth}
    \includegraphics[width=\textwidth]{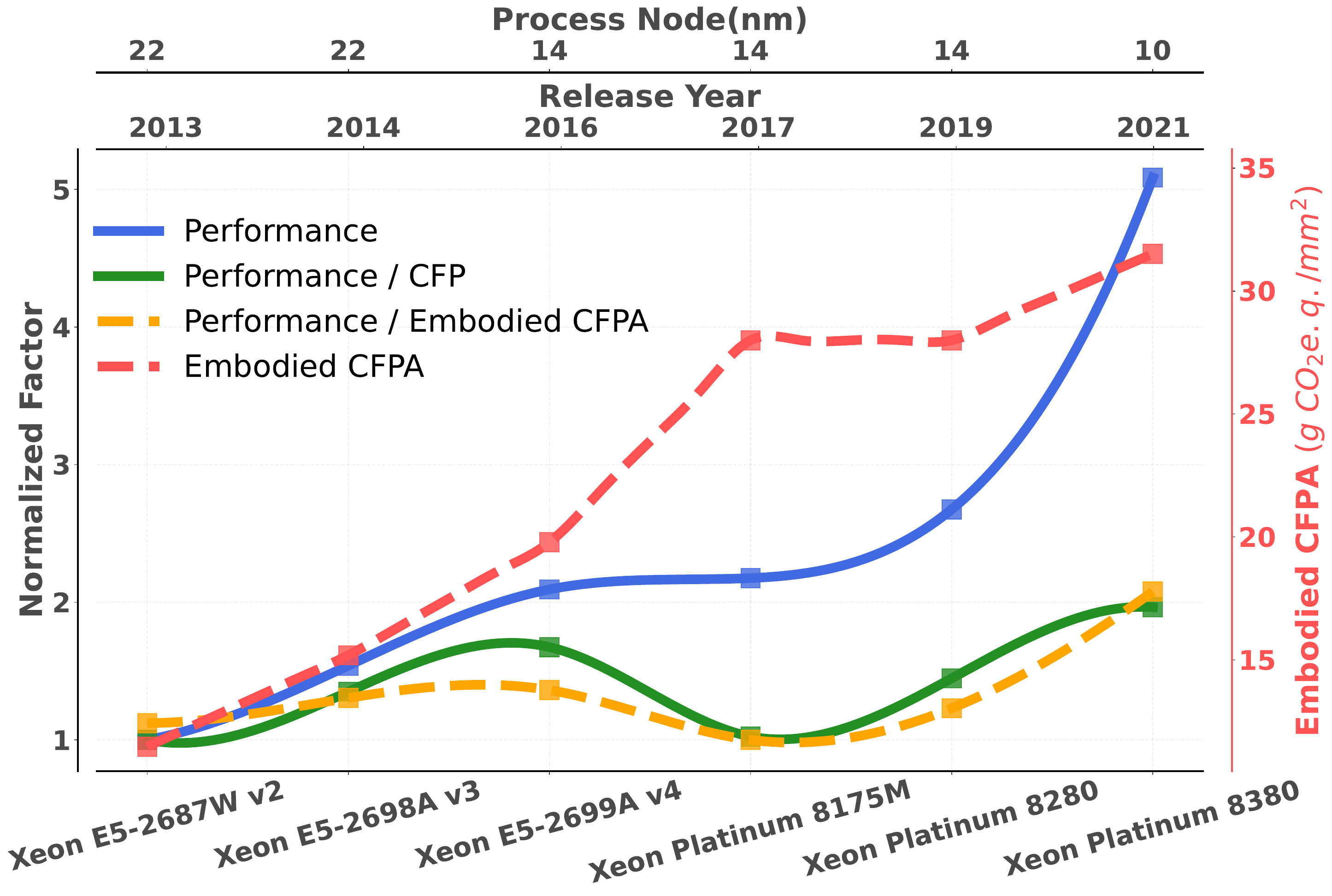}
    \label{fig:server_CPU_performance_metrics}
\end{subfigure}
\vspace{-3mm}
\caption{Sustainability trends in flagship desktop (top) and datacenter (bottom) CPUs from Intel}
\vspace{-4mm}
\label{fig:cpu_sustainability}
\end{figure*}

The two plots on the left in Figure. \ref{fig:cpu_sustainability}
show various metrics related to performance and sustainability for Intel's flagship desktop and datacenter CPUs in the last decade, respectively.
OCFP consistently dominates the overall CFP in both series while the proportion of ECFP has a steady growing trend in datacenter CPUs.
All CPUs show similar trends in process node evolution, transitioning from 22nm to 14nm, and eventually to 10nm. In desktop CPUs, each process node shrink leads to a significant reduction in die area and an increase in total CFP. Additionally, the die area gradually increases within the same process node, while total CFP slightly fluctuates. In contrast, the chip area of datacenter CPUs increases steadily with a slight dip at the end. Both series demonstrate a reduction in total CFP within the same process node, regardless of changes in chip area, shown by Core i7-6950X and Xeon Platinum 8280 at 14nm. This proves that while a CPU's total CFP decrease is primarily driven by process node shrinkage, the iterative architectural optimizations within the same process node can effectively enhance processor sustainability.

The right two plots in Figure \ref{fig:cpu_sustainability}
show various performance and sustainability metrics for Intel flagship CPUs in the last decade, respectively.
Overall, the trend for datacenter CPUs is smoother evidencing a more consistent architectural design philosophy. In contrast, desktop CPUs follow greater variability in processor architecture exploration within the same process node leading to violent fluctuations. Specifically, within 14nm process node, Core i7-6960X has both higher performance and ECFPA than Core i7-8700X but leads to a lower Performance/CFP. This demonstrates that architectural design has a great impact on the performance-sustainability tradeoff and it is feasible to design a processor with high performance and low CFP. Subsequently, from 14nm process node all the way to 10nm, Intel kept the performance/ECFPA, performance/CFP steadily increased, despite the rise in both ECFP and ECFPA, which aligns with the idea of sustainable design.

An opposite trend is observed in datacenter CPUs, where overall performance continues to increase steadily without achieving a good sustainability-performance balance. As process nodes advance, the performance gains do not convert into better sustainability. Both Performance/ECFPA and Performance/CFP show little improvement failing to keep pace with absolute performance. Especially, performance gained a boost at 10nm, but Performance/ECFP and Performance/ECFPA have limited increases. Generally, for datacenter CPUs, the increase in absolute performance is usually accompanied by a significant rise in TDP. Although the share of OCFP is decreasing, the aggressive power consumption continues to limit datacenter processors' sustainability.
\vspace{-5mm}

\begin{figure}[t]
    \centering
    \includegraphics[width=0.9\linewidth]{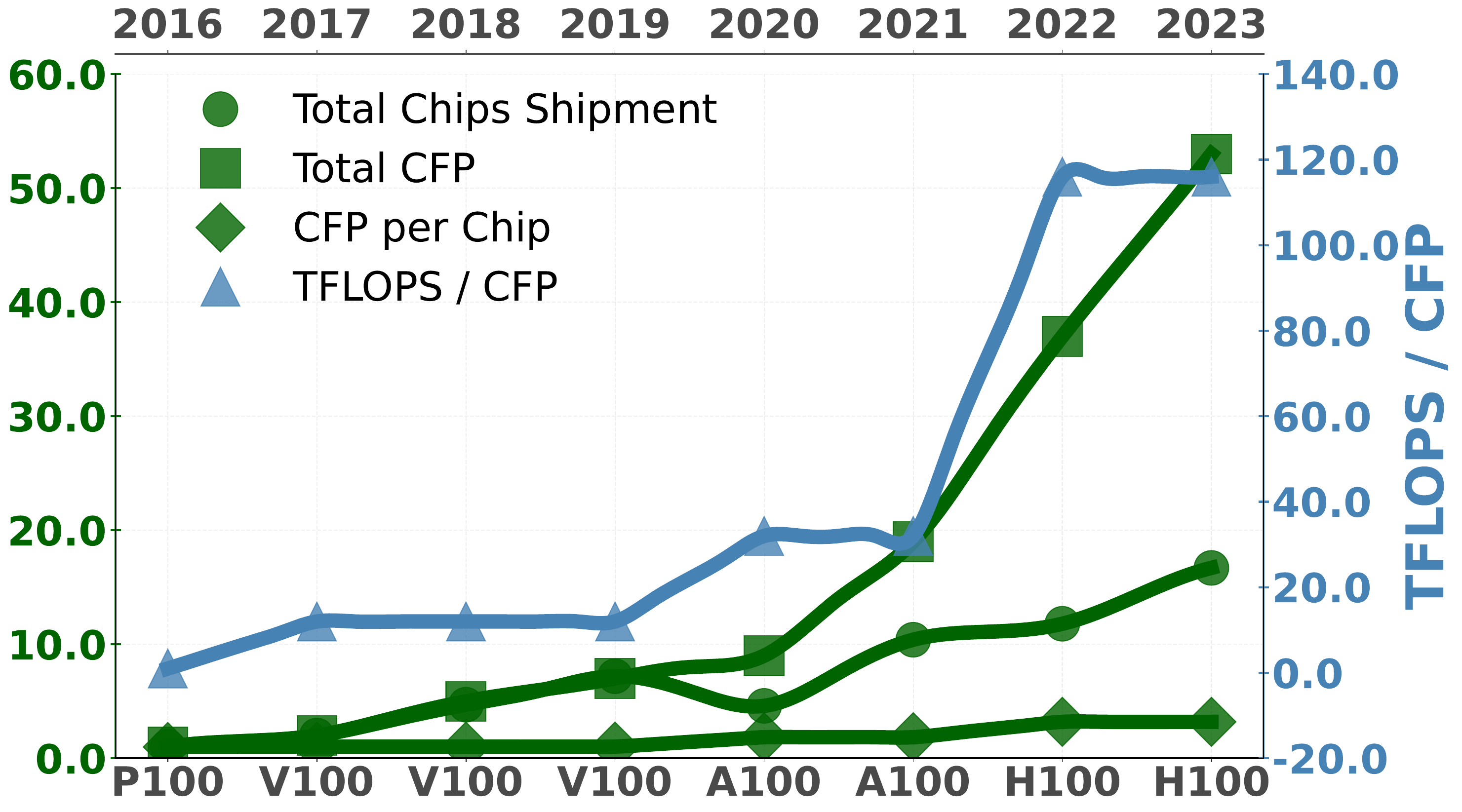}
    \caption{The AI boom greatly increased GPU shipments, raising total CFP despite the increased TFLOPS/CFP.}
    \vspace{-5mm}
    \label{fig:GPU_cfp_trend}
\end{figure}

\subsection{How has the AI boom impacted CFP?}
With the surging demand for high-performance GPUs in data centers due to rapidly expanding AI models, estimating datacenter GPU CFP growth provides insights into the impact of this AI boom. To this end, we first estimate the GPU shipments using NVIDIA's Datacenter Business Group annual financial reports \cite{nvidia_financial_reports}. We assume that the group’s revenue comes solely from the latest flagship GPU sales with a 75\% profit margin while being sold at the highest price. Additionally, we use the peak performance from datasheets as the performance estimator, since there is no publicly available single-chip AI benchmarking across generations of GPUs (e.g. MLPerf \cite{mlperf} scores are unavailable for older GPUs like the P100, and scores are for large multi-GPU systems). Since datacenter GPUs are usually sold as multi-GPU systems the actual single chip price is lower than the released price, meaning the true number of GPU shipments is likely much larger, making this a conservative estimation.

Fig.\ref{fig:GPU_cfp_trend} shows that although performance efficiency (TFLOPS/CFP) has improved dramatically, reaching 120X of the 2016 baseline, the total CFP per chip has not increased significantly.

The addition of domain-specific accelerator blocks and features (e.g., Tensor Cores in V100, structured sparsity and reduced precisions in A100, and transformer engine in H100) improved performance significantly at a minor increase in per-chip CFP, indicating a good sustainability trend. Except for a slight decline in 2020, GPU shipments have increased annually, driven in part by the recent surge in training and deployment of LLMs. This sharp increase has led to an explosive rise in overall $CO_2$ emissions, now exceeding 50X of the 2016 baseline.
This is a call for the community to design efficient chips and algorithms to stem the increase in CFP to keep the AI revolution healthy and green.
\footnote{The 5nm process node of H100 is actually not modeled in ECO-CHIP, but scaled based on existing process node metrics.}
\vspace{-3mm}

\begin{figure}[t]
    \centering
    \includegraphics[width=0.85\linewidth]{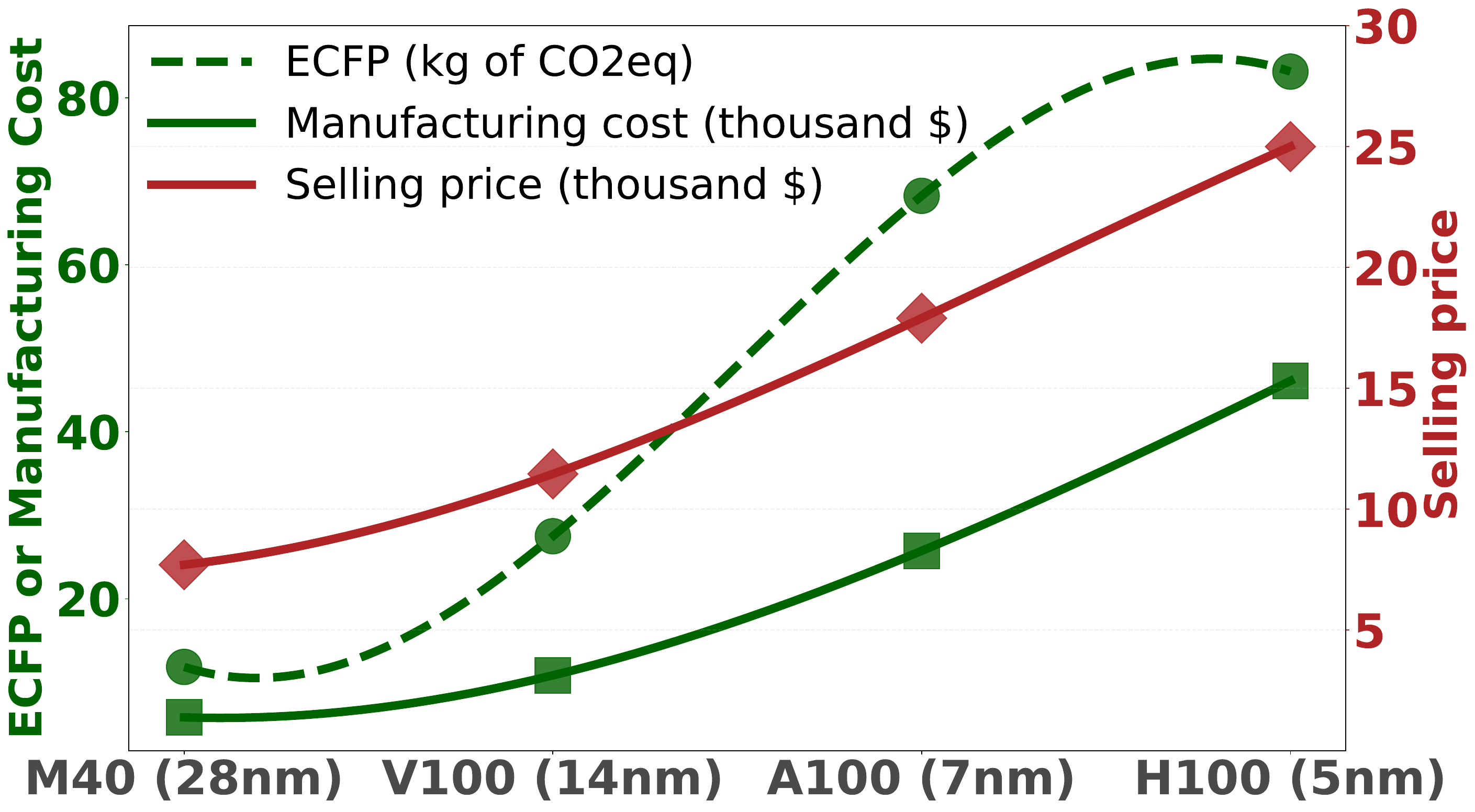}
    \caption{Manufacturing cost or selling price are not proper proxies of ECFP.}
    \label{fig:cost-vs-ecfp}
\end{figure}

\subsection{Is manufacturing cost (\$) a proxy for ECFP? }
With the growing importance of measuring and reducing carbon emissions, prior work has proposed using cost as a proxy for carbon emissions.
E.g., models like EIO-LCA \cite{cost-is-incorrect} estimate carbon emissions based on the economic cost of electronics, generally converting component costs into carbon emissions. 
Using our dataset, we analyze the validity of this consideration.
Figure~\ref{fig:cost-vs-ecfp} illustrates the variation in manufacturing cost along with ECFP. The divergence between these two trends becomes increasingly apparent as technology advances towards smaller nodes. This clearly demonstrates that manufacturing cost does not correlate with carbon. 
The plot also shows the selling price, another cost metric that could be used instead of manufacturing cost. However, we observe that it is not correlated with embodied carbon.
Since the selling price includes inflation, demand, supply chain, and profit margins, it is not an ideal metric for such considerations.

\begin{figure}
    \centering
    \includegraphics[width=0.8\linewidth]{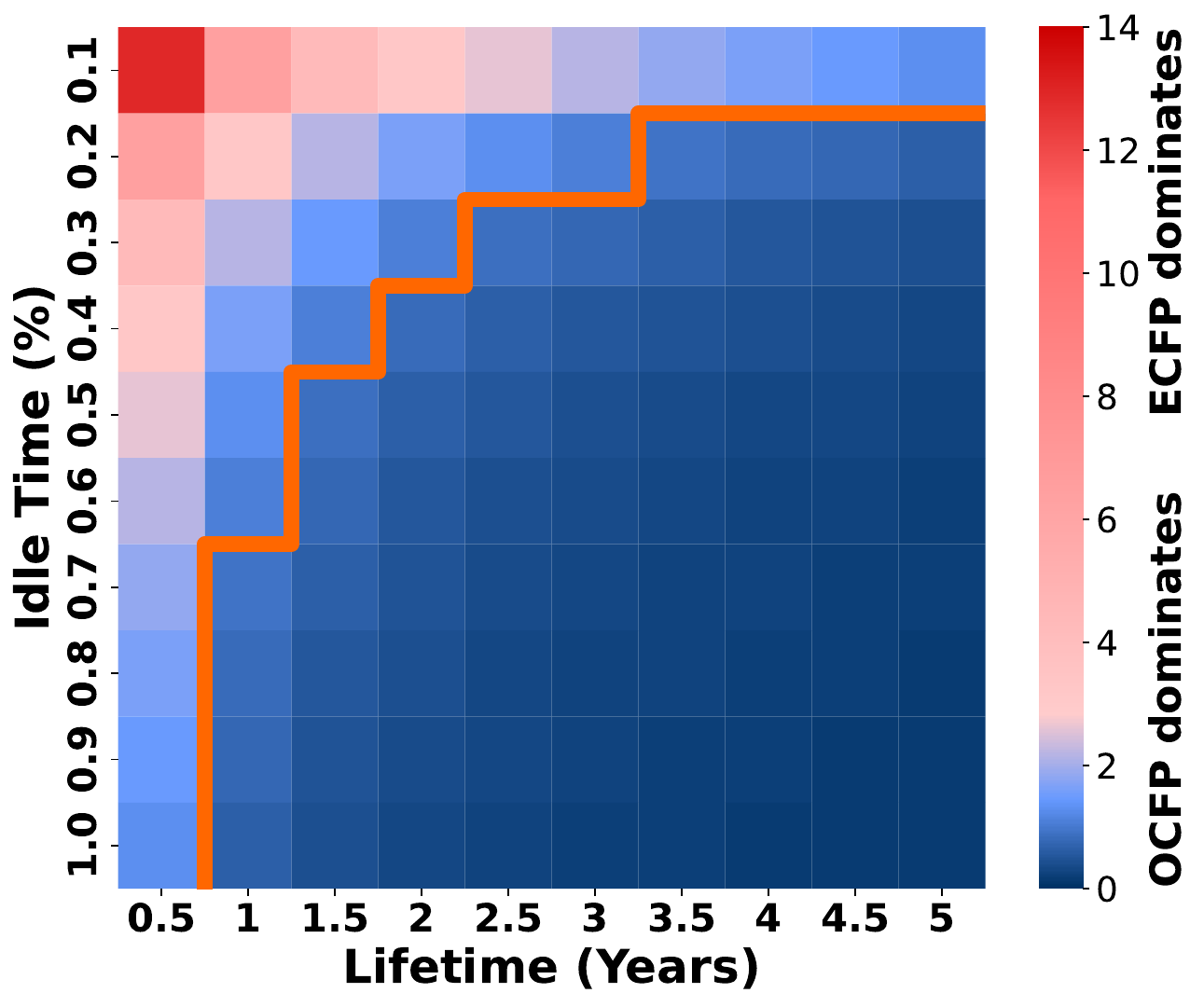}
    \caption{Ratio of ECFP to OCFP for NVIDIA A100-SXM across varying lifetime and idle time. The orange line indicates where ECFP exceeds or falls below OCFP.}
    \vspace{-5mm}
    \label{fig:emb-ope-ratio}
\end{figure}

\subsection{How much must processor lifetime increase by to effectively amortize ECFP?}

Modern computing systems are often replaced quickly for better performance. However, extending usage helps amortize the processor’s ECFP, a sunk cost, as examined in this study. Additionally, processors experience idle periods ~\cite{gpu-idle-time, cpu-idle-time}, causing OCFP to vary significantly with utilization.

Using the NVIDIA A100-SXM GPU as an example, we analyze the impact of processor lifetime and idle time, with lifetime ranging from 0.5 to 5 years and idle time varying from 0\% (always active) to 90\% (inactive 90\% of the time). Figure~\ref{fig:emb-ope-ratio} illustrates the ECFP-to-OCFP ratio, where regions above the orange line indicate that embodied emissions outweigh operational emissions. To effectively amortize ECFP, users must either extend the device’s lifetime or reduce idle time to shift below the orange line, ensuring OCFP dominance. For instance, with 70\% idle time, running the A100-SXM for more than two years enables effective ECFP amortization.

\begin{figure}
    \centering
    \begin{subfigure}{1\linewidth}
        \centering
        \includegraphics[width=\linewidth]{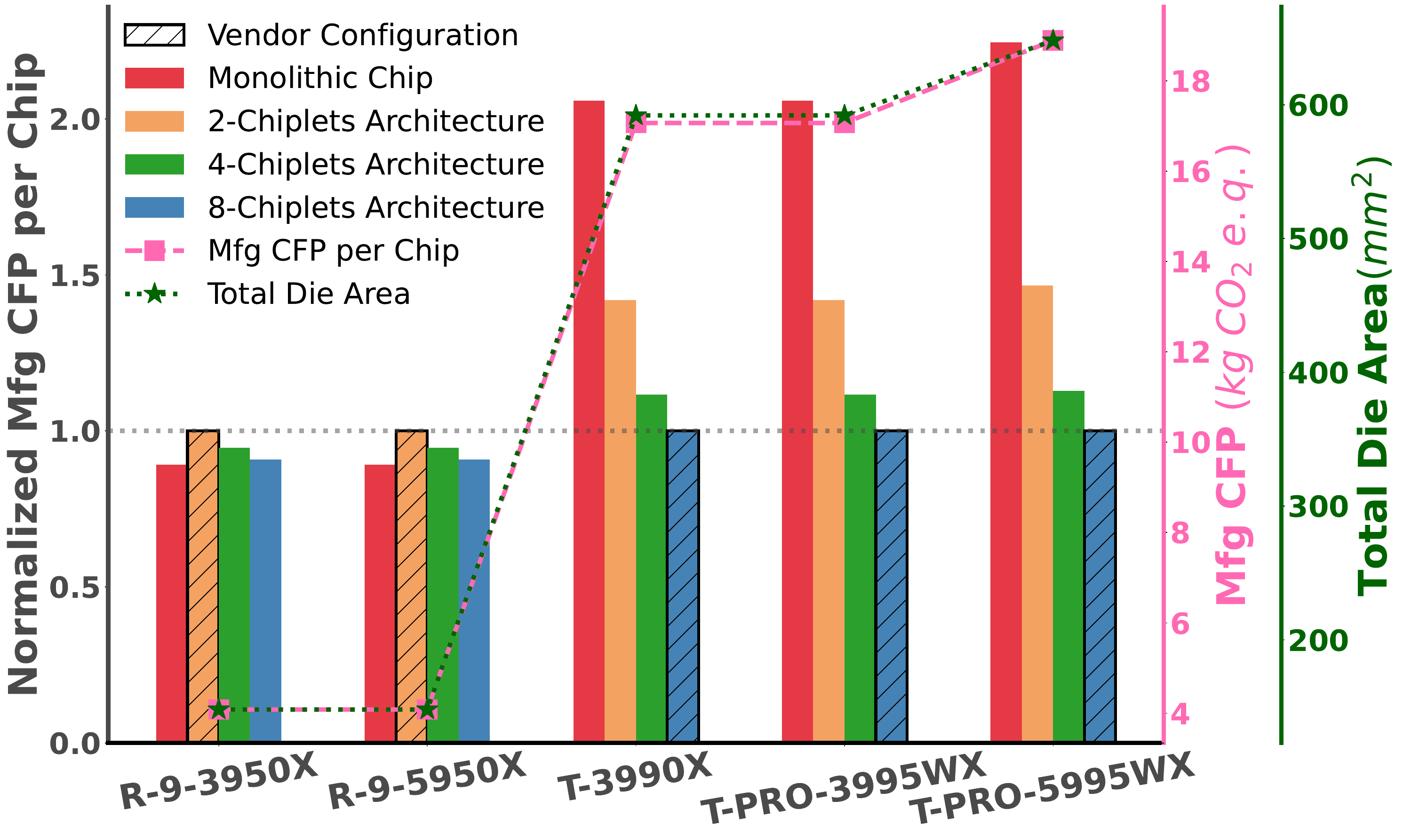}
    \end{subfigure}
    \hfill
    \begin{subfigure}{0.8\linewidth}
        \centering
        \includegraphics[width=\linewidth]{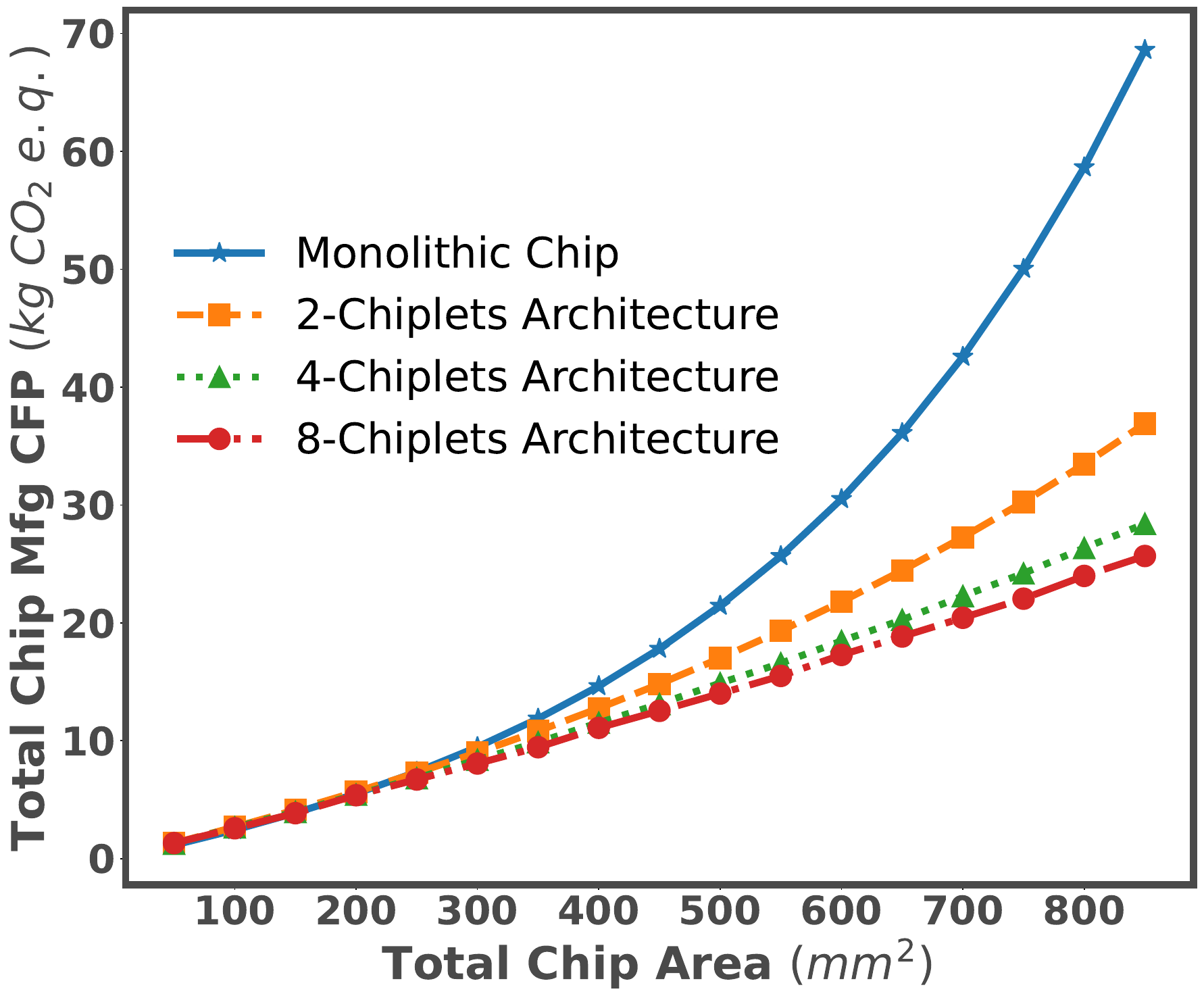}
    \end{subfigure}
    \caption{Manufacturing CFP of AMD flagship chiple CPUs (top) and monolithic, 2, 4, and 8 Chiplets architectures with total chip area scaled from 50 to 850 mm\(^2\) (bottom) all at 7nm. R=Ryzen, T=Threadripper. While more chiplets reduce CFP as area increases, monolithic architecture remains the most sustainable for chips under 200$mm^2$}
    \label{fig:chiplet-combined}
    \vspace{-5mm}
\end{figure}

\subsection{Are chiplet-based processors always more sustainable than
monolithic processors?}
Chiplet architecture, introduced by AMD in 2017, offers a pathway to extending Moore’s Law, but raises sustainability questions regarding yield gains versus increased area and packaging overhead. To explore this, we analyze the manufacturing CFP of AMD’s flagship chiplet CPUs over the past five years, varying chiplet counts within a fixed total chip area. Assuming an even distribution of total chip area across all chiplets, manufactured using the same process node, we also examine how manufacturing CFP scales across different chiplet configurations to assess whether chiplet architecture consistently provides a more sustainable solution. 
The extra packaging overhead is already modeled in our framework.

Fig.\ref{fig:chiplet-combined} (top) shows the normalized manufacturing CFP of AMD chiplet CPUs to the vendor configuration in different chiplet quantities. Our analysis reveals that chiplet architecture is not always the most sustainable solution, particularly for small to medium chips. Specifically, the monolithic design yields the lowest manufacturing CFP for the Ryzen 9 series, which has the smallest chip area, while the Threadripper series benefits the most from an increased chiplet count. Additionally, for the Ryzen 9 3950X and Ryzen 9 5950X, the yield improvement from a 2-chiplet design does not sufficiently offset the overhead of a higher chiplet count, making the monolithic architecture a more sustainable choice.

Figure~\ref{fig:chiplet-combined} (bottom) shows the trend of total manufacturing CFP across various chiplet configurations over a wide range of chip areas. The optimal architecture shifts with chip area, converging to designs with higher chiplet counts beyond a certain threshold. For chips smaller than $200mm^2$, monolithic architecture remains the most sustainable, while for those exceeding $300mm^2$ a higher chiplet count leads to better sustainability.

In practice, sustainability is not the primary driver for adopting chiplet-based designs; instead, they offer advantages in efficiency, testability, and manufacturing cost. Despite higher carbon emissions, performance remains the priority. For the Ryzen 9 series, while the 2-chiplet architecture is the least sustainable, its manufacturing CFP is comparable to other configurations. Its superior performance and efficiency, combined with lower manufacturing cost and better testability than higher chiplet-count designs, make it a balanced trade-off.

\vspace{-3mm}
\section{Potential Impact of CarbonSet}
\noindent
CarbonSet benchmarks processor sustainability by analyzing CFP trends over time, aiding carbon reduction and sustainable design. We showcased this through trend analysis and case studies but note its broader applicability, including:
\begin{itemize}
    \item{\textit{Life cycle assessment (LCA)}: Our dataset enables companies to estimate device lifetime CFP for LCA, supporting corporate sustainability reports and eco-labeling for consumer awareness.}
    \item {\textit{Optimizing designs and technology}: Historical data reveals which design aspects (e.g., manufacturing, materials, efficiency) have the highest carbon impact, guiding emissions-reducing optimizations for future processors.}
    \item {\textit{Environmental policy and standardization}:} Governments and industries can use this dataset to set carbon benchmarks, shape regulations, and incentivize low-carbon designs.
    \item {\textit{Education and carbon awareness}:} It can support academic and industrial research focused on environmental sustainability. It can also impact consumer purchasing decisions and help create a carbon-aware mindset.
\end{itemize}


\section{Conclusion} \label{sec:conclusion}
In this paper, we introduced CarboSet, a dataset designed for analyzing processor sustainability trends and enhancing sustainability-driven design evaluation. By leveraging this dataset, we examine key aspects of processor sustainability, including overall design trends over the past decade, the impact of the AI boom, the reliability of approximating ECFP using manufacturing costs, the required operational lifetime for effective ECFP amortization, and the sustainability implications of chiplet architectures. 
CarbonSet aims to guide environmentally conscious decisions in processor design, manufacturing, and lifecycle management.

\bibliographystyle{ACM-Reference-Format}
\bibliography{refs}


\begin{thebibliography}{27}


\ifx \showCODEN    \undefined \def \showCODEN     #1{\unskip}     \fi
\ifx \showISBNx    \undefined \def \showISBNx     #1{\unskip}     \fi
\ifx \showISBNxiii \undefined \def \showISBNxiii  #1{\unskip}     \fi
\ifx \showISSN     \undefined \def \showISSN      #1{\unskip}     \fi
\ifx \showLCCN     \undefined \def \showLCCN      #1{\unskip}     \fi
\ifx \shownote     \undefined \def \shownote      #1{#1}          \fi
\ifx \showarticletitle \undefined \def \showarticletitle #1{#1}   \fi
\ifx \showURL      \undefined \def \showURL       {\relax}        \fi
\providecommand\bibfield[2]{#2}
\providecommand\bibinfo[2]{#2}
\providecommand\natexlab[1]{#1}
\providecommand\showeprint[2][]{arXiv:#2}

\bibitem[Berger et~al\mbox{.}({[n.\,d.]})]%
        {david-brooks-sigarch-blog}
\bibfield{author}{\bibinfo{person}{D.~S. Berger}, \bibinfo{person}{D. Brooks}, \bibinfo{person}{F. Kazhamiaka}, \bibinfo{person}{M.~D. Hill}, \bibinfo{person}{R. Bianchini}, \bibinfo{person}{C.-J. Wu}, \bibinfo{person}{K. Strauss}, \bibinfo{person}{K. Frost}, \bibinfo{person}{J. Wang}, \bibinfo{person}{K. Martins}, \bibinfo{person}{S. Gillett}, \bibinfo{person}{E. Choukse}, \bibinfo{person}{D. Ernst}, \bibinfo{person}{R. Fonseca}, \bibinfo{person}{K. Lio}, \bibinfo{person}{B. Narayanasetty}, \bibinfo{person}{P. Patel}, \bibinfo{person}{C. Irvene}, \bibinfo{person}{A. Sriraman}, \bibinfo{person}{G. Porter}, \bibinfo{person}{A. Jones}, \bibinfo{person}{U. Gupta}, \bibinfo{person}{B. Acun-Uyan}, \bibinfo{person}{K. Hazelwood}, {and} \bibinfo{person}{D. Carmean}.} \bibinfo{year}{[n.\,d.]}\natexlab{}.
\newblock \bibinfo{title}{{Embodied Carbon is Important}}.
\newblock
\urldef\tempurl%
\url{https://www.sigarch.org/reducing-embodied-carbon-is-important}
\showURL{%
\tempurl}
\newblock
\shownote{Accessed: 2024-11-19}.


\bibitem[Bhagavathula et~al\mbox{.}(2024)]%
        {bhagavathula:understanding:2024}
\bibfield{author}{\bibinfo{person}{Anvita Bhagavathula}, \bibinfo{person}{Leo Han}, {and} \bibinfo{person}{Udit Gupta}.} \bibinfo{year}{2024}\natexlab{}.
\newblock \showarticletitle{Understanding the {{Implications}} of {{Uncertainty}} in {{Embodied Carbon Models}} for {{Sustainable Computing}}}. In \bibinfo{booktitle}{\emph{HotCarbon}}.
\newblock


\bibitem[Chien({[n.\,d.]})]%
        {andrew-chien-sigarch-blog}
\bibfield{author}{\bibinfo{person}{Andrew~A. Chien}.} \bibinfo{year}{[n.\,d.]}\natexlab{}.
\newblock \bibinfo{title}{{Why Embodied Carbon is a poor Architecture Design metric, and Operational Carbon remains an important Problem}}.
\newblock
\urldef\tempurl%
\url{https://www.sigarch.org/why-embodied-carbon-is-a-poor-architecture-design-metric-and-operational-carbon-remains-an-important-problem/}
\showURL{%
\tempurl}
\newblock
\shownote{Accessed: 2024-11-19}.


\bibitem[Choppali~Sudarshan et~al\mbox{.}(2024)]%
        {greenFPGA}
\bibfield{author}{\bibinfo{person}{Chetan Choppali~Sudarshan}, \bibinfo{person}{Aman Arora}, {and} \bibinfo{person}{Vidya~A Chhabria}.} \bibinfo{year}{2024}\natexlab{}.
\newblock \showarticletitle{{GreenFPGA: Evaluating FPGAs as Environmentally Sustainable Computing Solutions}}. In \bibinfo{booktitle}{\emph{Proceedings of the 61st ACM/IEEE Design Automation Conference}}. \bibinfo{pages}{1--6}.
\newblock


\bibitem[Corporation(2024)]%
        {nvidia_financial_reports}
\bibfield{author}{\bibinfo{person}{NVIDIA Corporation}.} \bibinfo{year}{2024}\natexlab{}.
\newblock \bibinfo{title}{Financial Reports}.
\newblock
\urldef\tempurl%
\url{https://investor.nvidia.com/Home/default.aspx}
\showURL{%
\tempurl}
\newblock
\shownote{Accessed: 2024-11-19}.


\bibitem[Eeckhout(2022)]%
        {first_order}
\bibfield{author}{\bibinfo{person}{Lieven Eeckhout}.} \bibinfo{year}{2022}\natexlab{}.
\newblock \showarticletitle{{A First-Order Model to Assess Computer Architecture Sustainability}}.
\newblock \bibinfo{journal}{\emph{IEEE Computer Architecture Letters}} \bibinfo{volume}{21}, \bibinfo{number}{2} (\bibinfo{year}{2022}), \bibinfo{pages}{137--140}.
\newblock
\href{https://doi.org/10.1109/LCA.2022.3217366}{doi:\nolinkurl{10.1109/LCA.2022.3217366}}


\bibitem[{Exxact}(2022)]%
        {gpu-idle-time}
\bibfield{author}{\bibinfo{person}{{Exxact}}.} \bibinfo{year}{2022}\natexlab{}.
\newblock \bibinfo{title}{{{Run:ai: You've got Idle GPUs. We Guarantee It}}}.
\newblock
\urldef\tempurl%
\url{https://www.exxactcorp.com/blog/Deep-Learning/run-ai-you-ve-got-idle-gpus-we-guarantee-it}
\showURL{%
\tempurl}
\newblock
\shownote{Accessed: 2024-11-19}.


\bibitem[{Geekbench}(2024)]%
        {Geekbench}
\bibfield{author}{\bibinfo{person}{{Geekbench}}.} \bibinfo{year}{2024}\natexlab{}.
\newblock \bibinfo{title}{{Geekbench-Processor Benchmark}}.
\newblock
\urldef\tempurl%
\url{https://browser.geekbench.com/processor-benchmarks}
\showURL{%
\tempurl}
\newblock
\shownote{Accessed: 2024-11-19}.


\bibitem[{Georgia, Butler}(2023)]%
        {cpu-idle-time}
\bibfield{author}{\bibinfo{person}{{Georgia, Butler}}.} \bibinfo{year}{2023}\natexlab{}.
\newblock \bibinfo{title}{{{Study: Only 13\% of provisioned CPUs and 20\% of memory utilized in cloud computing}}}.
\newblock
\urldef\tempurl%
\url{https://www.datacenterdynamics.com/en/news/only-13-of-provisioned-cpus-and-20-of-memory-utilized-in-cloud-computing-report/}
\showURL{%
\tempurl}
\newblock
\shownote{Accessed: 2024-11-19}.


\bibitem[Gupta et~al\mbox{.}(2022a)]%
        {ACT}
\bibfield{author}{\bibinfo{person}{Udit Gupta}, \bibinfo{person}{Mariam Elgamal}, \bibinfo{person}{Gage Hills}, \bibinfo{person}{Gu-Yeon Wei}, \bibinfo{person}{Hsien-Hsin~S. Lee}, \bibinfo{person}{David Brooks}, {and} \bibinfo{person}{Carole-Jean Wu}.} \bibinfo{year}{2022}\natexlab{a}.
\newblock \showarticletitle{{ACT}: designing sustainable computer systems with an architectural carbon modeling tool}. In \bibinfo{booktitle}{\emph{Proceedings of the 49th {Annual} {International} {Symposium} on {Computer} {Architecture}}}. \bibinfo{publisher}{ACM}, \bibinfo{address}{New York New York}, \bibinfo{pages}{784--799}.
\newblock
\showISBNx{978-1-4503-8610-4}
\href{https://doi.org/10.1145/3470496.3527408}{doi:\nolinkurl{10.1145/3470496.3527408}}


\bibitem[Gupta et~al\mbox{.}(2022b)]%
        {chasing_carbon}
\bibfield{author}{\bibinfo{person}{Udit Gupta}, \bibinfo{person}{Young~Geun Kim}, \bibinfo{person}{Sylvia Lee}, \bibinfo{person}{Jordan Tse}, \bibinfo{person}{Hsien-Hsin~S. Lee}, \bibinfo{person}{Gu-Yeon Wei}, \bibinfo{person}{David Brooks}, {and} \bibinfo{person}{Carole-Jean Wu}.} \bibinfo{year}{2022}\natexlab{b}.
\newblock \showarticletitle{{Chasing Carbon: The Elusive Environmental Footprint of Computing}}.
\newblock \bibinfo{journal}{\emph{IEEE Micro}} \bibinfo{volume}{42}, \bibinfo{number}{4} (\bibinfo{year}{2022}), \bibinfo{pages}{37--47}.
\newblock
\href{https://doi.org/10.1109/MM.2022.3163226}{doi:\nolinkurl{10.1109/MM.2022.3163226}}


\bibitem[Henderson et~al\mbox{.}(2020)]%
        {ml_carbon}
\bibfield{author}{\bibinfo{person}{Peter Henderson}, \bibinfo{person}{Jieru Hu}, \bibinfo{person}{Joshua Romoff}, \bibinfo{person}{Emma Brunskill}, \bibinfo{person}{Dan Jurafsky}, {and} \bibinfo{person}{Joelle Pineau}.} \bibinfo{year}{2020}\natexlab{}.
\newblock \showarticletitle{{Towards the systematic reporting of the energy and carbon footprints of machine learning}}.
\newblock \bibinfo{journal}{\emph{J. Mach. Learn. Res.}} \bibinfo{volume}{21}, \bibinfo{number}{1}, Article \bibinfo{articleno}{248} (\bibinfo{date}{Jan.} \bibinfo{year}{2020}), \bibinfo{numpages}{43}~pages.
\newblock
\showISSN{1532-4435}


\bibitem[Hendrickson et~al\mbox{.}(1998)]%
        {cost-is-incorrect}
\bibfield{author}{\bibinfo{person}{Chris Hendrickson}, \bibinfo{person}{Arpad Horvath}, \bibinfo{person}{Satish Joshi}, \bibinfo{person}{Octavio Juarez}, \bibinfo{person}{Lester Lave}, \bibinfo{person}{H~Scott Matthews}, \bibinfo{person}{Francis~C McMichael}, {and} \bibinfo{person}{Elisa Cobas-Flores}.} \bibinfo{year}{1998}\natexlab{}.
\newblock \showarticletitle{{Economic Input-Output-Based Life Cycle Assessment (EIO-LCA)}}.
\newblock \bibinfo{journal}{\emph{mental}} (\bibinfo{year}{1998}).
\newblock


\bibitem[Hoefflinger(2012)]%
        {itrs}
\bibfield{author}{\bibinfo{person}{Bernd Hoefflinger}.} \bibinfo{year}{2012}\natexlab{}.
\newblock \bibinfo{booktitle}{\emph{ITRS: The International Technology Roadmap for Semiconductors}}.
\newblock \bibinfo{publisher}{Springer Berlin Heidelberg}, \bibinfo{address}{Berlin, Heidelberg}, \bibinfo{pages}{161--174}.
\newblock
\showISBNx{978-3-642-23096-7}
\href{https://doi.org/{10.1007/978-3-642-23096-7_7}}{doi:\nolinkurl{{10.1007/978-3-642-23096-7_7}}}


\bibitem[{Ian Cutress}(2020)]%
        {def-den}
\bibfield{author}{\bibinfo{person}{{Ian Cutress}}.} \bibinfo{year}{2020}\natexlab{}.
\newblock \bibinfo{title}{{Better Yield on 5nm than 7nm: TSMC Update on Defect Rates for N5}}.
\newblock
\urldef\tempurl%
\url{https://www.anandtech.com/show/16028/better-yield-on-5nm-than-7nm-tsmc-update-on-defect-rates-for-n5}
\showURL{%
\tempurl}
\newblock
\shownote{Accessed: 2024-11-19}.


\bibitem[{J.Hu, C.C Sudarshan, V.A. Chhabria, A.Arora}(2025)]%
        {carbonset_github}
\bibfield{author}{\bibinfo{person}{{J.Hu, C.C Sudarshan, V.A. Chhabria, A.Arora}}.} \bibinfo{year}{2025}\natexlab{}.
\newblock \bibinfo{title}{{CarbonSet GitHub}}.
\newblock
\urldef\tempurl%
\url{https://github.com/advent-lab/CarbonSet}
\showURL{%
\tempurl}
\newblock
\shownote{Accessed: 2025-4-19}.


\bibitem[{Marie Garcia Bardon}({[n.\,d.]})]%
        {gpa-imec}
\bibfield{author}{\bibinfo{person}{{Marie Garcia Bardon}}.} \bibinfo{year}{[n.\,d.]}\natexlab{}.
\newblock \bibinfo{title}{{{The environmental footprint of logic CMOS technologies}}}.
\newblock
\urldef\tempurl%
\url{https://www.imec-int.com/en/articles/environmental-footprint-logic-cmos-technologies}
\showURL{%
\tempurl}
\newblock
\shownote{Accessed: 2024-11-19}.


\bibitem[Mattson et~al\mbox{.}(2019)]%
        {mlperf}
\bibfield{author}{\bibinfo{person}{Peter Mattson}, \bibinfo{person}{Christine Cheng}, \bibinfo{person}{Cody Coleman}, \bibinfo{person}{Greg Diamos}, \bibinfo{person}{Paulius Micikevicius}, \bibinfo{person}{David~A. Patterson}, \bibinfo{person}{Hanlin Tang}, \bibinfo{person}{Gu{-}Yeon Wei}, \bibinfo{person}{Peter Bailis}, \bibinfo{person}{Victor Bittorf}, \bibinfo{person}{David Brooks}, \bibinfo{person}{Dehao Chen}, \bibinfo{person}{Debojyoti Dutta}, \bibinfo{person}{Udit Gupta}, \bibinfo{person}{Kim~M. Hazelwood}, \bibinfo{person}{Andrew Hock}, \bibinfo{person}{Xinyuan Huang}, \bibinfo{person}{Bill Jia}, \bibinfo{person}{Daniel Kang}, \bibinfo{person}{David Kanter}, \bibinfo{person}{Naveen Kumar}, \bibinfo{person}{Jeffery Liao}, \bibinfo{person}{Guokai Ma}, \bibinfo{person}{Deepak Narayanan}, \bibinfo{person}{Tayo Oguntebi}, \bibinfo{person}{Gennady Pekhimenko}, \bibinfo{person}{Lillian Pentecost}, \bibinfo{person}{Vijay~Janapa Reddi}, \bibinfo{person}{Taylor Robie}, \bibinfo{person}{Tom~St. John},
  \bibinfo{person}{Carole{-}Jean Wu}, \bibinfo{person}{Lingjie Xu}, \bibinfo{person}{Cliff Young}, {and} \bibinfo{person}{Matei Zaharia}.} \bibinfo{year}{2019}\natexlab{}.
\newblock \showarticletitle{{MLPerf Training Benchmark}}.
\newblock \bibinfo{journal}{\emph{CoRR}}  \bibinfo{volume}{abs/1910.01500} (\bibinfo{year}{2019}).
\newblock
\showeprint[arXiv]{1910.01500}
\urldef\tempurl%
\url{http://arxiv.org/abs/1910.01500}
\showURL{%
\tempurl}


\bibitem[{Our World Data}(2024)]%
        {ci-world}
\bibfield{author}{\bibinfo{person}{{Our World Data}}.} \bibinfo{year}{2024}\natexlab{}.
\newblock \bibinfo{title}{{Carbon intensity of electricity generation}}.
\newblock
\urldef\tempurl%
\url{{https://tinyurl.com/carbon-intensity-world}}
\showURL{%
\tempurl}
\newblock
\shownote{Accessed: 2024-11-19}.


\bibitem[{PassMark}(2024)]%
        {PassMark}
\bibfield{author}{\bibinfo{person}{{PassMark}}.} \bibinfo{year}{2024}\natexlab{}.
\newblock \bibinfo{title}{{PassMark-Processor Performance Test}}.
\newblock
\urldef\tempurl%
\url{https://www.passmark.com/products/performancetest/}
\showURL{%
\tempurl}
\newblock
\shownote{Accessed: 2024-11-19}.


\bibitem[{Standard Performance Evaluation Corporation}(2024)]%
        {spec}
\bibfield{author}{\bibinfo{person}{{Standard Performance Evaluation Corporation}}.} \bibinfo{year}{2024}\natexlab{}.
\newblock \bibinfo{title}{{SPEC Benchmarks and Tools}}.
\newblock
\urldef\tempurl%
\url{https://www.spec.org/benchmarks.html}
\showURL{%
\tempurl}
\newblock
\shownote{Accessed: 2024-11-19}.


\bibitem[Sudarshan et~al\mbox{.}(2024a)]%
        {metrics}
\bibfield{author}{\bibinfo{person}{C.~C. Sudarshan}, \bibinfo{person}{A. Arora}, {and} \bibinfo{person}{V.~A Chhabria}.} \bibinfo{year}{2024}\natexlab{a}.
\newblock \showarticletitle{{Beyond the Surface: The Necessity of Detailed Metrics in Corporate Sustainability}}. In \bibinfo{booktitle}{\emph{Proc. of the 15th International Green and Sustainable Computing Conference}} (Austin, US) \emph{(\bibinfo{series}{IGSCC '24})}.
\newblock


\bibitem[Sudarshan et~al\mbox{.}(2024b)]%
        {eco-chip}
\bibfield{author}{\bibinfo{person}{Chetan~Choppali Sudarshan}, \bibinfo{person}{Nikhil Matkar}, \bibinfo{person}{Sarma Vrudhula}, \bibinfo{person}{Sachin~S. Sapatnekar}, {and} \bibinfo{person}{Vidya~A. Chhabria}.} \bibinfo{year}{2024}\natexlab{b}.
\newblock \showarticletitle{{ ECO-CHIP: Estimation of Carbon Footprint of Chiplet-based Architectures for Sustainable VLSI }}. In \bibinfo{booktitle}{\emph{2024 IEEE International Symposium on High-Performance Computer Architecture (HPCA)}}. \bibinfo{publisher}{IEEE Computer Society}, \bibinfo{address}{Los Alamitos, CA, USA}, \bibinfo{pages}{671--685}.
\newblock
\href{https://doi.org/10.1109/HPCA57654.2024.00058}{doi:\nolinkurl{10.1109/HPCA57654.2024.00058}}


\bibitem[Sun et~al\mbox{.}(2019)]%
        {dataset}
\bibfield{author}{\bibinfo{person}{Yifan Sun}, \bibinfo{person}{Nicolas~Bohm Agostini}, \bibinfo{person}{Shi Dong}, {and} \bibinfo{person}{David~R. Kaeli}.} \bibinfo{year}{2019}\natexlab{}.
\newblock \showarticletitle{{Summarizing {CPU} and {GPU} Design Trends with Product Data}}.
\newblock \bibinfo{journal}{\emph{CoRR}}  \bibinfo{volume}{abs/1911.11313} (\bibinfo{year}{2019}).
\newblock
\showeprint[arXiv]{1911.11313}
\urldef\tempurl%
\url{http://arxiv.org/abs/1911.11313}
\showURL{%
\tempurl}


\bibitem[{TSMC}(2019)]%
        {epa-tsmc}
\bibfield{author}{\bibinfo{person}{{TSMC}}.} \bibinfo{year}{2019}\natexlab{}.
\newblock \bibinfo{title}{{Corporate Socual Responsibility Report 2019}}.
\newblock
\urldef\tempurl%
\url{https://esg.tsmc.com/download/csr/2019-csr-report/english/pdf/e-all.pdf}
\showURL{%
\tempurl}
\newblock
\shownote{Accessed: 2024-11-19}.


\bibitem[Vahdat et~al\mbox{.}(2024)]%
        {metrics_acm}
\bibfield{author}{\bibinfo{person}{Amin Vahdat}, \bibinfo{person}{Xiaoyu Ma}, {and} \bibinfo{person}{David Patterson}.} \bibinfo{year}{2024}\natexlab{}.
\newblock \showarticletitle{{New Computer Evaluation Metrics for a Changing World}}.
\newblock \bibinfo{journal}{\emph{Commun. ACM}} \bibinfo{volume}{67}, \bibinfo{number}{10} (\bibinfo{date}{Sept.} \bibinfo{year}{2024}), \bibinfo{pages}{31–33}.
\newblock
\showISSN{0001-0782}
\href{https://doi.org/10.1145/3637867}{doi:\nolinkurl{10.1145/3637867}}


\bibitem[Wu et~al\mbox{.}(2021)]%
        {sustainableAI}
\bibfield{author}{\bibinfo{person}{Carole{-}Jean Wu}, \bibinfo{person}{Ramya Raghavendra}, \bibinfo{person}{Udit Gupta}, \bibinfo{person}{Bilge Acun}, \bibinfo{person}{Newsha Ardalani}, \bibinfo{person}{Kiwan Maeng}, \bibinfo{person}{Gloria Chang}, \bibinfo{person}{Fiona~Aga Behram}, \bibinfo{person}{James Huang}, \bibinfo{person}{Charles Bai}, \bibinfo{person}{Michael Gschwind}, \bibinfo{person}{Anurag Gupta}, \bibinfo{person}{Myle Ott}, \bibinfo{person}{Anastasia Melnikov}, \bibinfo{person}{Salvatore Candido}, \bibinfo{person}{David Brooks}, \bibinfo{person}{Geeta Chauhan}, \bibinfo{person}{Benjamin Lee}, \bibinfo{person}{Hsien{-}Hsin~S. Lee}, \bibinfo{person}{Bugra Akyildiz}, \bibinfo{person}{Maximilian Balandat}, \bibinfo{person}{Joe Spisak}, \bibinfo{person}{Ravi Jain}, \bibinfo{person}{Mike Rabbat}, {and} \bibinfo{person}{Kim~M. Hazelwood}.} \bibinfo{year}{2021}\natexlab{}.
\newblock \showarticletitle{{Sustainable {AI:} Environmental Implications, Challenges and Opportunities}}.
\newblock \bibinfo{journal}{\emph{CoRR}}  \bibinfo{volume}{abs/2111.00364} (\bibinfo{year}{2021}).
\newblock
\showeprint[arXiv]{2111.00364}
\urldef\tempurl%
\url{https://arxiv.org/abs/2111.00364}
\showURL{%
\tempurl}


\end{thebibliography}

\end{document}